\documentclass[11pt]{article}

\RequirePackage[colorlinks,citecolor=blue,urlcolor=blue]{hyperref}

\usepackage{enumerate}
\usepackage{multicol}
\usepackage{multirow}
\usepackage{adjustbox}
\usepackage{url} 
\usepackage{booktabs}
\usepackage{bm}
\usepackage{arxiv}
\usepackage{natbib}
\usepackage{url} 
\usepackage{amssymb}
\usepackage{amsmath}
\usepackage{lmodern}
\usepackage{mathrsfs}%
\usepackage{subcaption}
\usepackage[title]{appendix}%
\usepackage{xcolor}%
\usepackage{booktabs}%
\usepackage[ruled]{algorithm2e}
\usepackage{listings}
\usepackage{amsfonts}
\usepackage{amsthm}
\usepackage{mathtools}
\mathtoolsset{showonlyrefs=true}
\usepackage{multirow}
\usepackage{float}
\usepackage{adjustbox}

\newcommand{\bolds}[1]{\boldsymbol{#1}}

\newcommand{\calH}{{\cal H}}

\newcommand{\calS}{{\cal S}}

\newcommand{\ba}{\bolds{a}}

\newcommand{\bb}{\bolds{b}}
\newcommand{\bB}{\mathbf{B}}

\newcommand{\bbE}{\mathbb{E}}

\newcommand{\bbR}{\mathbb{R}}

\newcommand{\bu}{\bolds{u}}

\newcommand{\bx}{\bolds{x}}

\newcommand{\by}{\bolds{y}}
\newcommand{\bY}{\bolds{Y}}

\newcommand{\beps}{\bolds{\varepsilon}}
\newcommand{\boeta}{\bolds{\eta}}
\newcommand{\btheta}{\bolds{\theta}}

\newcommand{\bnu}{\bolds{\nu}}

\newcommand{\bOm}{\bolds{\Omega}}

\newcommand{\given}{\,|\,}
\newcommand{\lrnd}{\left(}
\newcommand{\rrnd}{\right)}
\newcommand{\lsq}{\left[}
\newcommand{\rsq}{\right]}
\newcommand{\lcur}{\left\lbrace}
\newcommand{\rcur}{\right\rbrace}
\renewcommand{\tilde}{\widetilde}

\title{Environmental Risk Assessment via Nonhomogeneous Hidden Semi-Markov Models with Penalized Vector Auto-Regression }

\author{
Marco Mingione\\
    \scriptsize{Dipartimento di Statistica, Informatica, Applicazioni ``G. Parenti”}\\
    \scriptsize{Università degli Studi di Firenze}\\
    \scriptsize{\texttt{marco.mingione@unifi.it}}\\
    \And
Pierfrancesco Alaimo Di Loro\\
    \scriptsize{Dipartimento GEPLI}\\
    \scriptsize{Libera Università Maria Ss. Assunta (LUMSA)}\\
    \scriptsize{\texttt{p.alaimodiloro@lumsa.it}}\\
\And
    Francesco Lagona\\
    \scriptsize{Dipartimento di Scienze Politiche}\\
    \scriptsize{Università degli Studi Roma Tre}\\
    \scriptsize{\texttt{francesco.lagona@uniroma3.it}}\\
     \And
Antonello Maruotti\\
    \scriptsize{Dipartimento GEPLI}\\
    \scriptsize{Libera Università Maria Ss. Assunta (LUMSA)}\\
    \scriptsize{\texttt{a.maruotti@lumsa.it}}\\
}
\begin{document}
\maketitle


\begin{abstract}
Motivated by the study of pollution trends in the city of Bergen, we introduce a flexible statistical framework for modeling multivariate air pollution data via a nonhomogeneous Hidden Semi-Markov Vector Auto-Regression. The hidden process captures unobserved environmental conditions, while the vector autoregressive structure accounts for temporal autocorrelation and cross-pollutant dependencies. The model further allows time-varying environmental conditions to influence both the average levels of pollutant concentrations and the duration of different transient states. Parameters are estimated via maximum likelihood using a tailored Expectation-Maximization (EM) algorithm, integrated with state-specific $\ell_1$ regularization to control overfitting and automatically select relevant temporal lags. The proposal is tested on simulated data under different scenarios and then applied to daily concentrations of nitrogens and particulate matter recorded in a urban area. Environmental risk is assessed by a Shapley value-based decomposition that attribute marginal risk contributions. This approach offers a comprehensive framework for multivariate environmental risk modeling, enabling better identification of high-pollution episodes and informing policy interventions.
\end{abstract}



\section{Introduction}
\label{sec:intro}
Identifying the sources and patterns of air-pollution is essential for planning policy interventions meant to preserve the environment and public health \citep{greven2011approach, shan2024methods}.
Traditional methods have largely focused on single-pollutant models \citep{liang2021modeling,mork2024heterogeneous}, however air pollutants do not occur in isolation \citep{finazzi2013model,cao2024integration,zhu2024review} but are emitted and dispersed jointly. They often share common sources and co-occur under specific environmental conditions, which influence how they interact through atmospheric chemistry, particularly during episodes of atmospheric instability when simultaneous surges in pollutant concentrations occur. This makes the environmental risk associated with air pollution a fundamentally multivariate phenomenon that exhibits complex mixtures of behaviors \citep{boaz2019multivariate,baragano2022multiple}. As such, neglecting their joint dynamics can hinder a comprehensive understanding of pollution patterns.  

To address these issues, air pollution modeling must move beyond univariate approaches and adopt multivariate statistical frameworks that reflect the true complexity of pollution mixtures \citep{maruotti2017dynamic, bouveyron2022co}. 
We develop a general and flexible modeling approach that reflects the empirical features observed in air quality measurements -- features that are not only specific to air pollution trends but also broadly relevant across other applications involving multivariate time series.
We assume that pollutant concentrations can be generated under multiple unobserved environmental conditions that can favor or limit the accumulation of air pollutants. These states can be inferred from the data via a time-varying mixture model, where the hidden state dynamics are governed by a semi-Markov process \citep{barbu2009semi, yu2015hidden, ruiz2022hidden}. Unlike standard Markov models, the semi-Markov formulation allows for sojourn times that are not necessarily geometrically distributed, enabling a more realistic representation of the persistence and duration of pollution episodes. Moreover, we allow for the inclusion of covariates' effects (e.g., meteorological conditions) in the sojourn distribution, introducing an additional layer of flexibility in how long different environmental states might last \citep{lagona2025nonhomogeneous,koslik2025hidden}. 
Given the hidden state, we assume that the observed multivariate process -- i.e. the pollutant concentrations -- can be described by a Gaussian Vector Auto-Regressive model (VAR) with state-dependent parameters \citep{hadj2024bayesian}. This formulation allows accounting for temporal autocorrelation, which is a prominent feature of atmospheric pollutant series due to accumulation and transport mechanisms, and the instantaneous cross-dependence, allowing it to vary across different environmental conditions. For instance, pollutants' accumulation and correlations can strengthen during high-pollution episodes and weaken under cleaner conditions. This enables the model to reflect short-term dependencies and Granger-causal relationships among pollutants, capturing both feedback dynamics and temporal clustering in their evolution.
Though the model is very flexible, the computational burden is not cumbersome. Parameters are estimated by an Expectation-Maximization (EM) algorithm with a regularized M-step. Regularization is obtained by integrating the M-Step with a $\ell_1$ penalty on the auto-regressive coefficients, considering a state-dependent penalty as in \cite{stadler2013penalized}. In the absence of appropriate asymptotic results, estimation uncertainty is quantified by parametric bootstrap.

The estimated model is subsequently employed to quantify the contribution of individual pollutants to overall environmental risk. Drawing on the extensive literature on financial risk assessment, we adapt these methodologies to the environmental domain, emphasizing the importance of capturing the multidimensional nature of environmental risk. To this end, we adopt the multivariate risk framework introduced by \cite{tobias2016covar} and extended to dynamic mixtures in \cite{bernardi2017multiple}. 
In particular, we implement a \textit{standardized} Shapley value-based risk attribution methodology to ensure an equitable decomposition of each pollutant’s risk.

The model is motivated by an analysis of daily pollutant concentration levels of NO, NO$_2$, PM$_1$, PM$_{2.5}$ and PM$_{10}$ at Danmarksplass, the busiest traffic intersection in Bergen. It has garnered media attention due to poor air quality during cold, dry periods with minimal wind. Approximately 40,000 vehicles traverse this intersection daily and the surrounding contains a significant proportion of old wooden houses heated by wood-burning stoves \citep{bergen_air_quality_2023}.
The results reveal notable features of the multivariate time series of pollutants, particularly the emergence of two distinct states characterized by low and high concentration levels. Beyond differences in average pollutant levels, these states also exhibit markedly different correlation structures and temporal dynamics. Specifically, we notice that in the absence of favorable meteorological conditions the high-pollution state may persist indefinitely.
In terms of risk contribution, the analysis identifies the nitrogens (NO and NO$_2$) and PM$_{2.5}$ and PM$_{10}$ as mutually reinforcing in terms of risk. By contrast, PM$_1$ exhibits a more autonomous behavior, with a seasonally varying relationship to PM$_{2.5}$.
This seasonal asymmetry may reflect the shifting composition of pollution sources across the year.

\section{Data description}



We focus on daily air pollutant concentrations of particulate matter across three size categories ($\mathrm{PM}_{10}$, $\mathrm{PM}_{2.5}$, and $\mathrm{PM}_{1}$), as well as nitrogen oxides (NO and $\mathrm{NO}_2$), recorded from January 1, 2020, to December 31, 2022. All concentrations are measured in micrograms per cubic meter ($\mu g/m^3$). By definition, particulate matter concentrations of increasing size are intrinsically dependent. Specifically, 
\text{PM}$_{10} =$ \text{PM}$_{2.5} +$ \text{PM}$^*_{10}$ and \text{PM}$_{2.5} =$ \text{PM}$_{1} +$ \text{PM}$^*_{2.5}$, 
where $\mathrm{PM}_{10}^*$ and $\mathrm{PM}_{2.5}^*$ represent residual particulate matter concentrations corresponding to particles sized between 2.5–10 and 1–2.5, respectively. This structure reflects a natural hierarchy in particulate matter composition: $\mathrm{PM}_1$ includes the smallest particles, and larger particle categories are constructed by sequentially adding residual mass. Consequently, any valid probabilistic modeling of the PM variables must respect the ordering constraint $\mathrm{PM}_{10} \geq \mathrm{PM}_{2.5} \geq \mathrm{PM}_{1}$.
To address and disentangle this inherent dependence structure, we work with the residual PM components. We retain $\mathrm{PM}_{1}$ in its original form and define the residual variables as \text{PM}$^*_{10}$ = \text{PM}$_{10}-$\text{PM}$_{2.5}$ and finally \text{PM}$^*_{2.5} = $\text{PM}$_{2.5}-$\text{PM}$_{1}$.

\begin{figure}
        \centering
        \begin{subfigure}[b]{.33\textwidth}
            \centering
            \includegraphics[width = .95\textwidth]{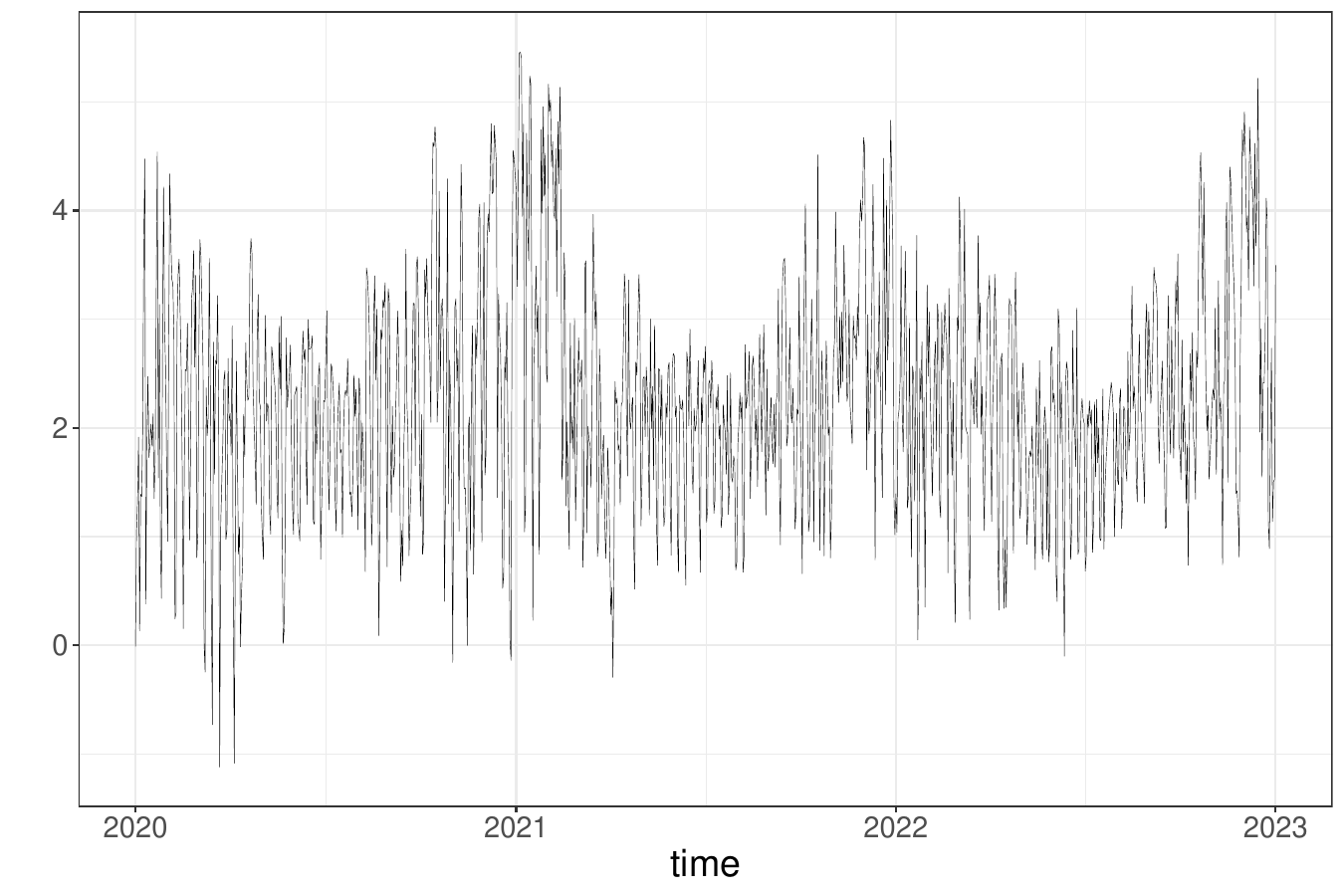}
            \caption*{\scriptsize NO}
        \end{subfigure}
        \begin{subfigure}[b]{.33\textwidth}
            \centering
            \includegraphics[width = .95\textwidth]{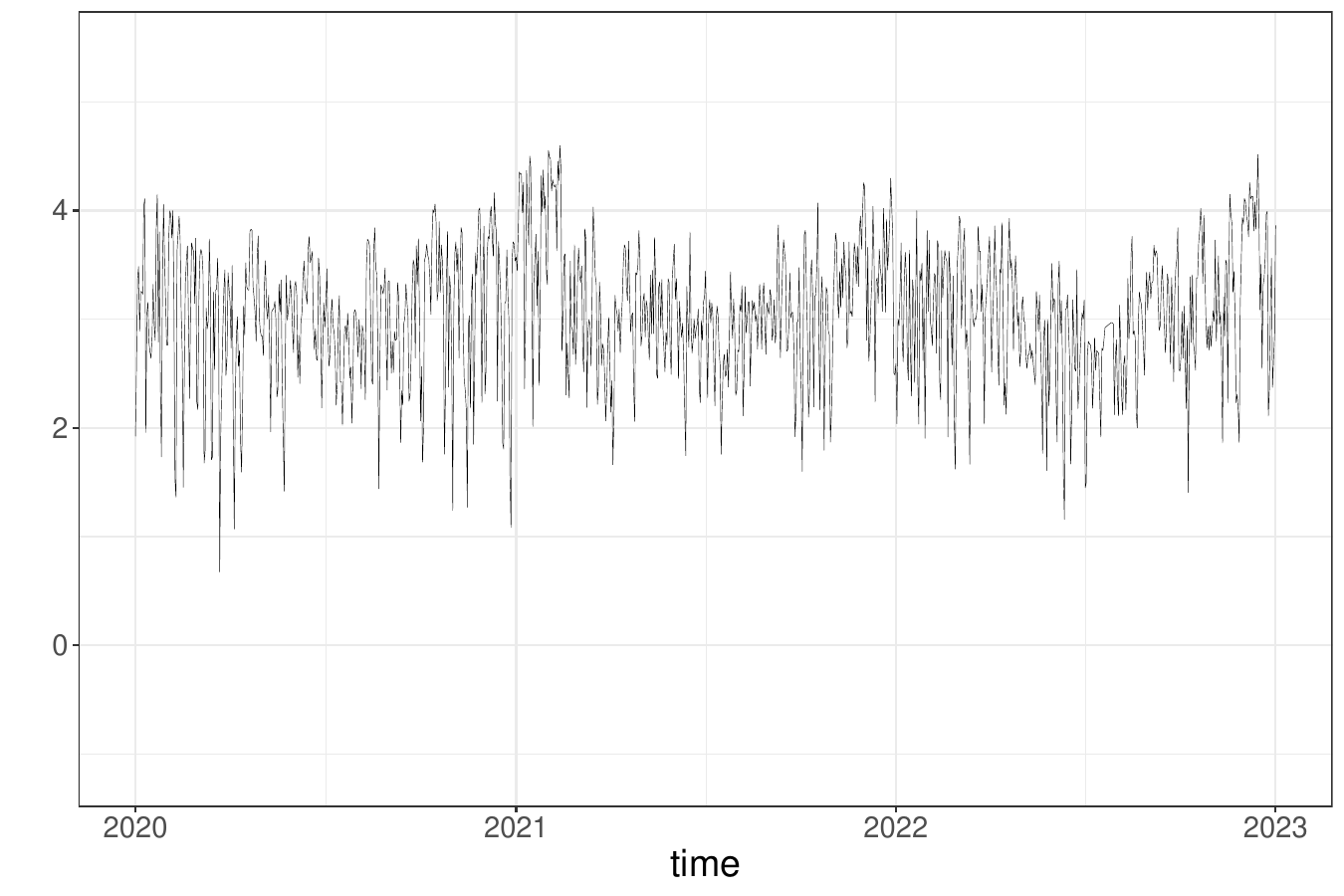}
            \caption*{\scriptsize NO$_2$}
        \end{subfigure}
        \\
        \begin{subfigure}[b]{.31\textwidth}
            \centering
            \includegraphics[width = .98\textwidth]{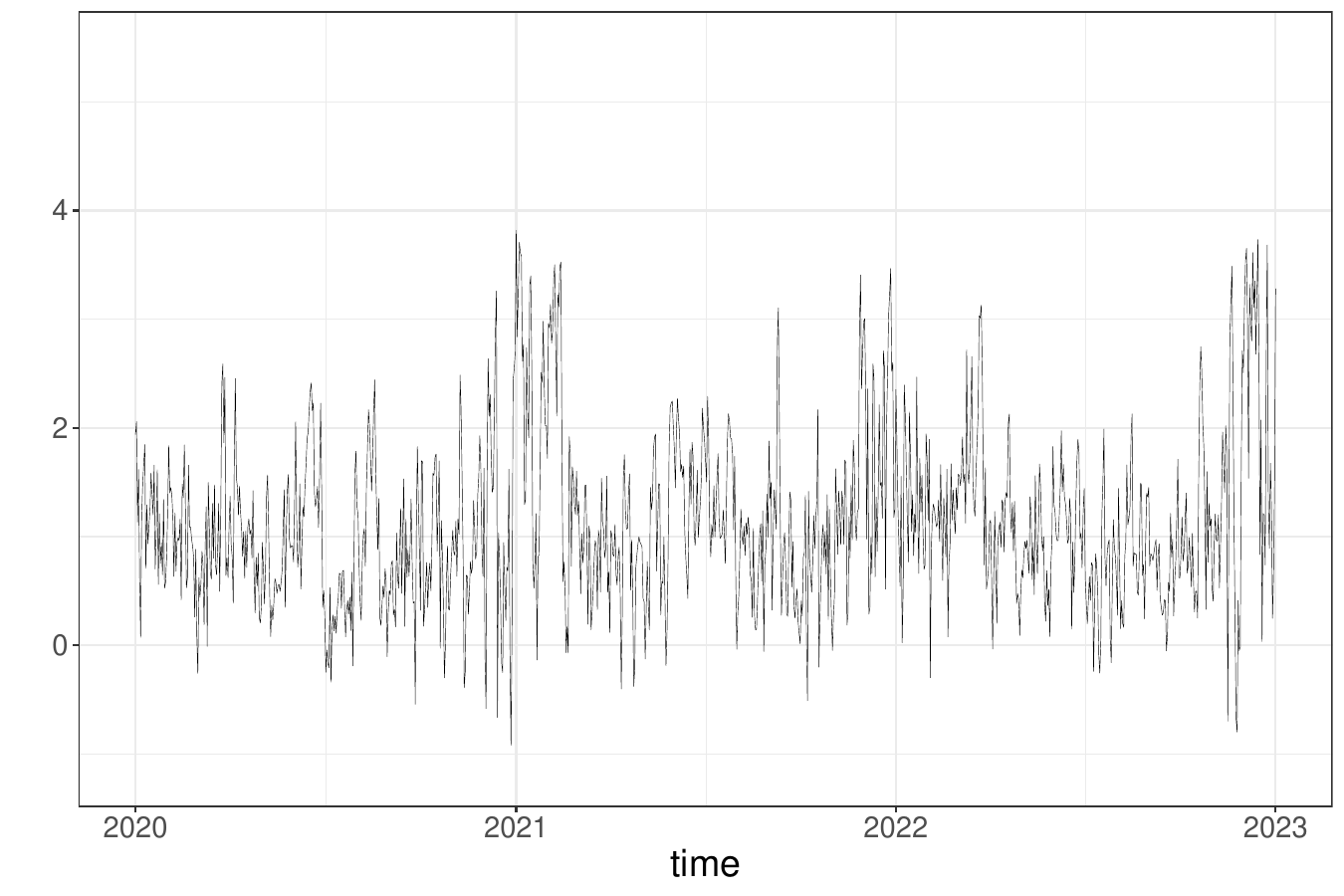}
            \caption*{\scriptsize PM$_1$}
        \end{subfigure}
        \begin{subfigure}[b]{.31\textwidth}
            \centering
            \includegraphics[width = .98\textwidth]{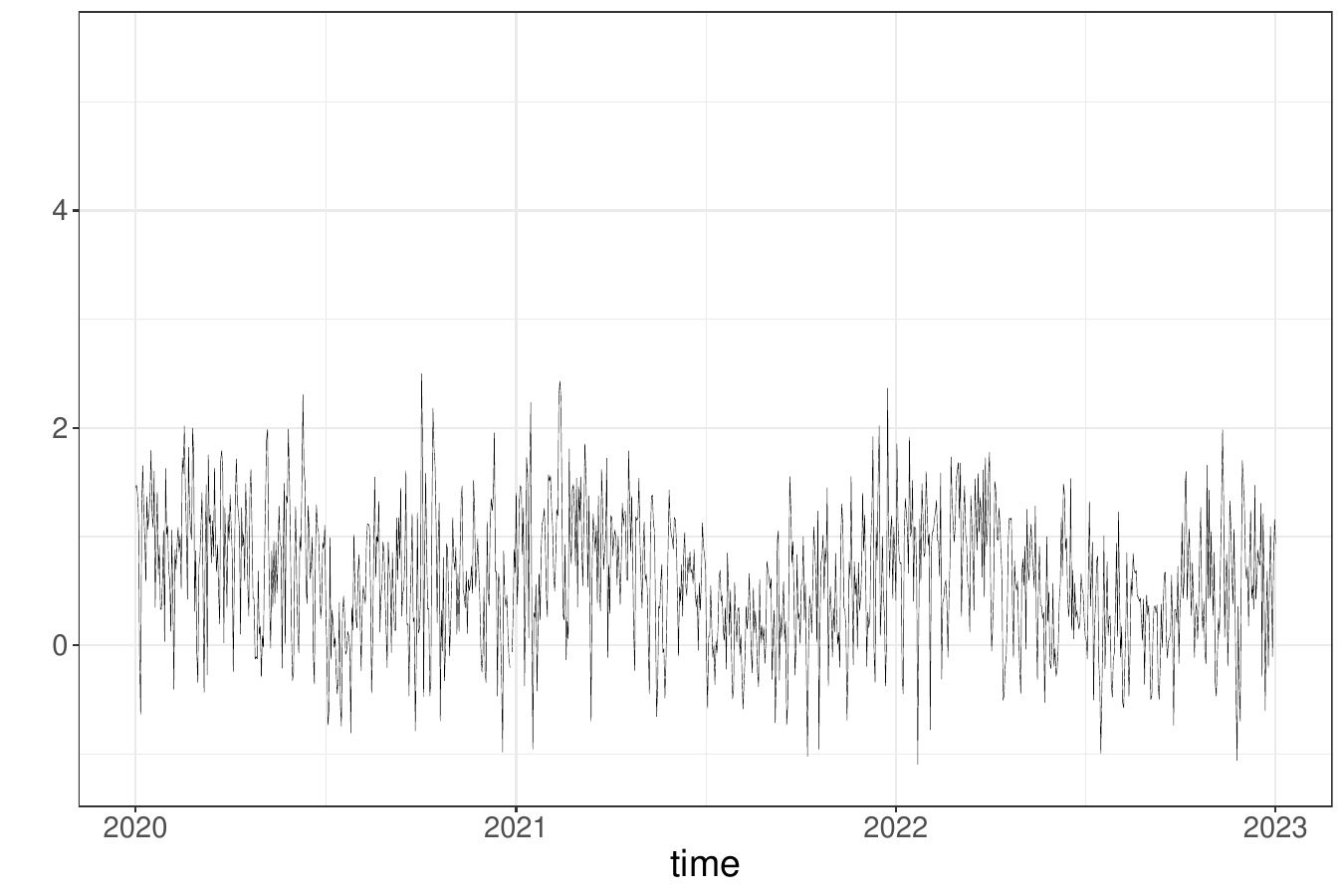}
            \caption*{\scriptsize PM$^*_{2.5}$}
        \end{subfigure}
        \begin{subfigure}[b]{.31\textwidth}
            \centering
            \includegraphics[width = .98\textwidth]{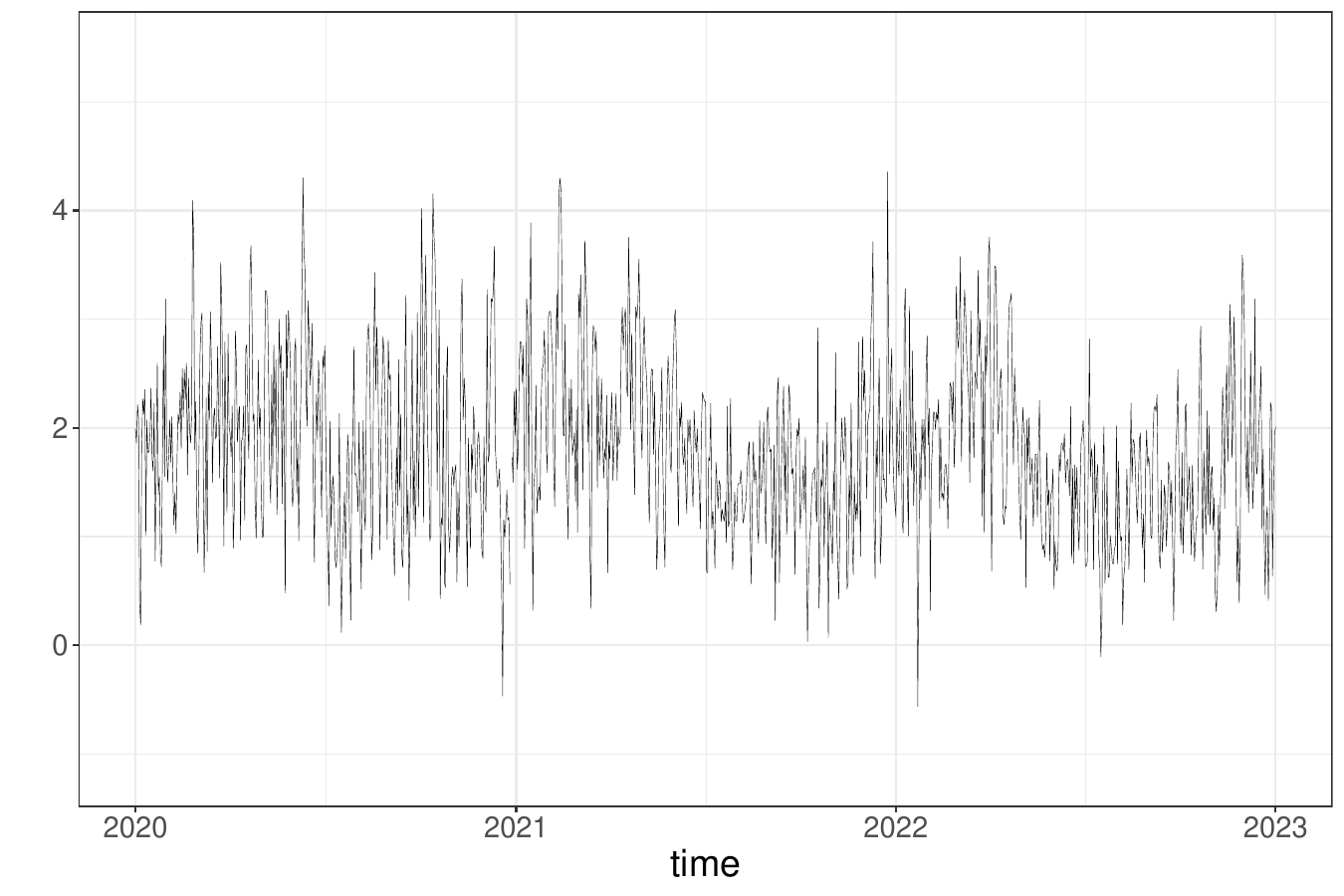}
            \caption*{\scriptsize PM$^*_{10}$}
        \end{subfigure}
        \caption{Observed time series (on the log-scale) of the five considered pollutants over the whole study period: (a) NO, (b) NO$_2$, (c) PM$_1$, (d) PM$^*_{2.5}$, (e) PM$^*_{10}$.}
        \label{fig:obsts}
    \end{figure}

Additionally, the raw concentration data are log-transformed to accommodate the multivariate Gaussian assumption of the emission density within the VAR model discussed in Section \ref{sec:meth_observation process}. The use of log-concentrations is well-established in the environmental literature \citep{liao2021multiple}, supported both by physical justifications for log-normality \citep{ott1990physical, andersson2021mechanisms} and by the interpretability of log-scale relationships as elasticities.
From this point onward, unless otherwise specified, the terms $\mathrm{PM}_{10}^*$, $\mathrm{PM}_{2.5}^*$, $\mathrm{PM}_{1}$, NO, and $\mathrm{NO}_2$ refer to the log-transformed pollutant concentrations.
Figure \ref{fig:obsts} presents the observed time series of daily pollutant concentrations. 
The data exhibit a broadly stationary pattern with yearly seasonality, while closely occurring spikes suggest a degree of temporal dependence. These patterns are further emphasized by the partial autocorrelation functions reported in the Supplementary Material, together with both the marginal and joint (pairwise) distributions of the log-transformed concentrations, supporting the Gaussian assumptions and the existence of possible heterogeneity.

\begin{figure}
        \centering
         \begin{subfigure}[b]{.31\textwidth}
            \centering
            \includegraphics[width = .95\textwidth]{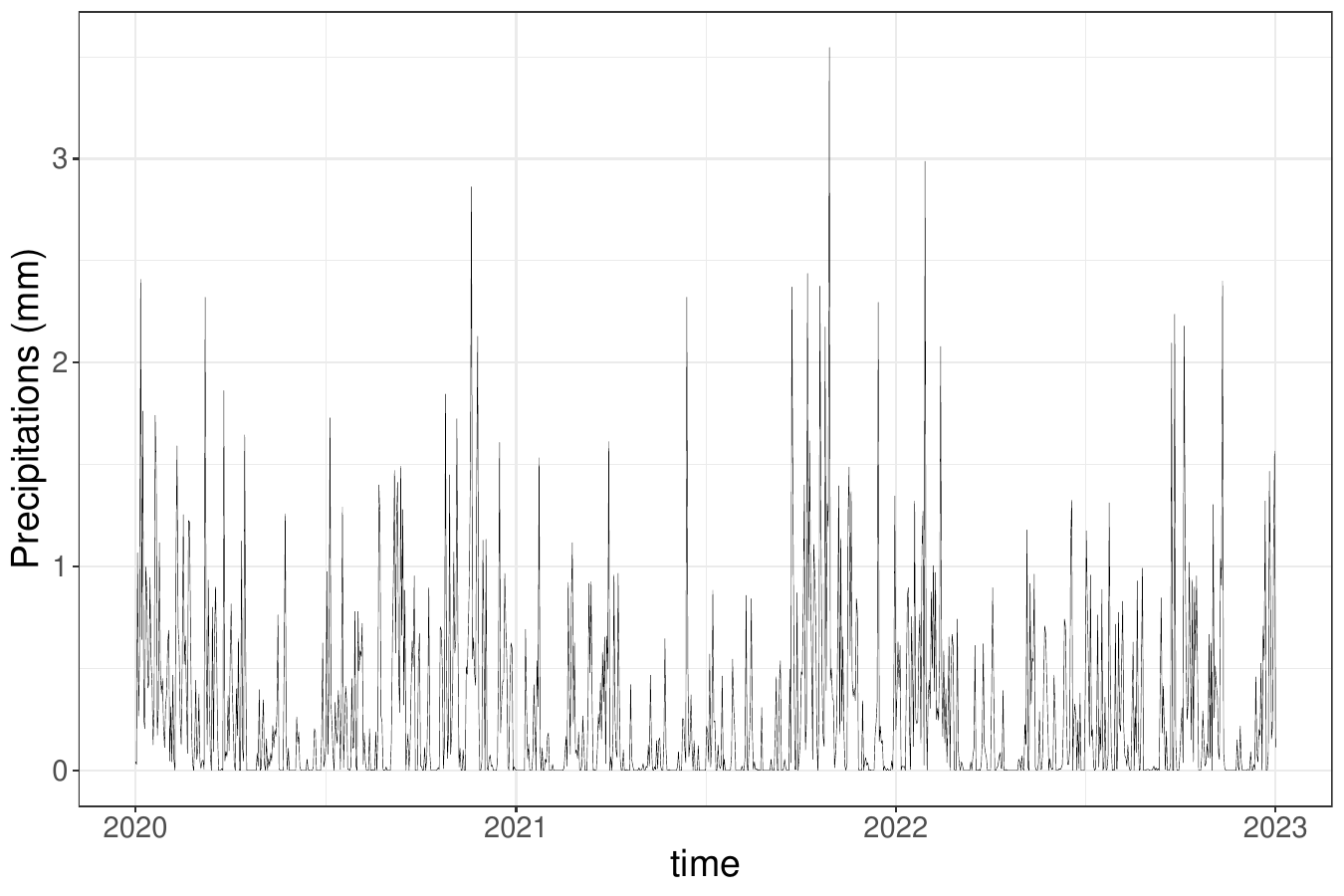}
        \end{subfigure}
        \begin{subfigure}[b]{.31\textwidth}
            \centering
            \includegraphics[width = .95\textwidth]{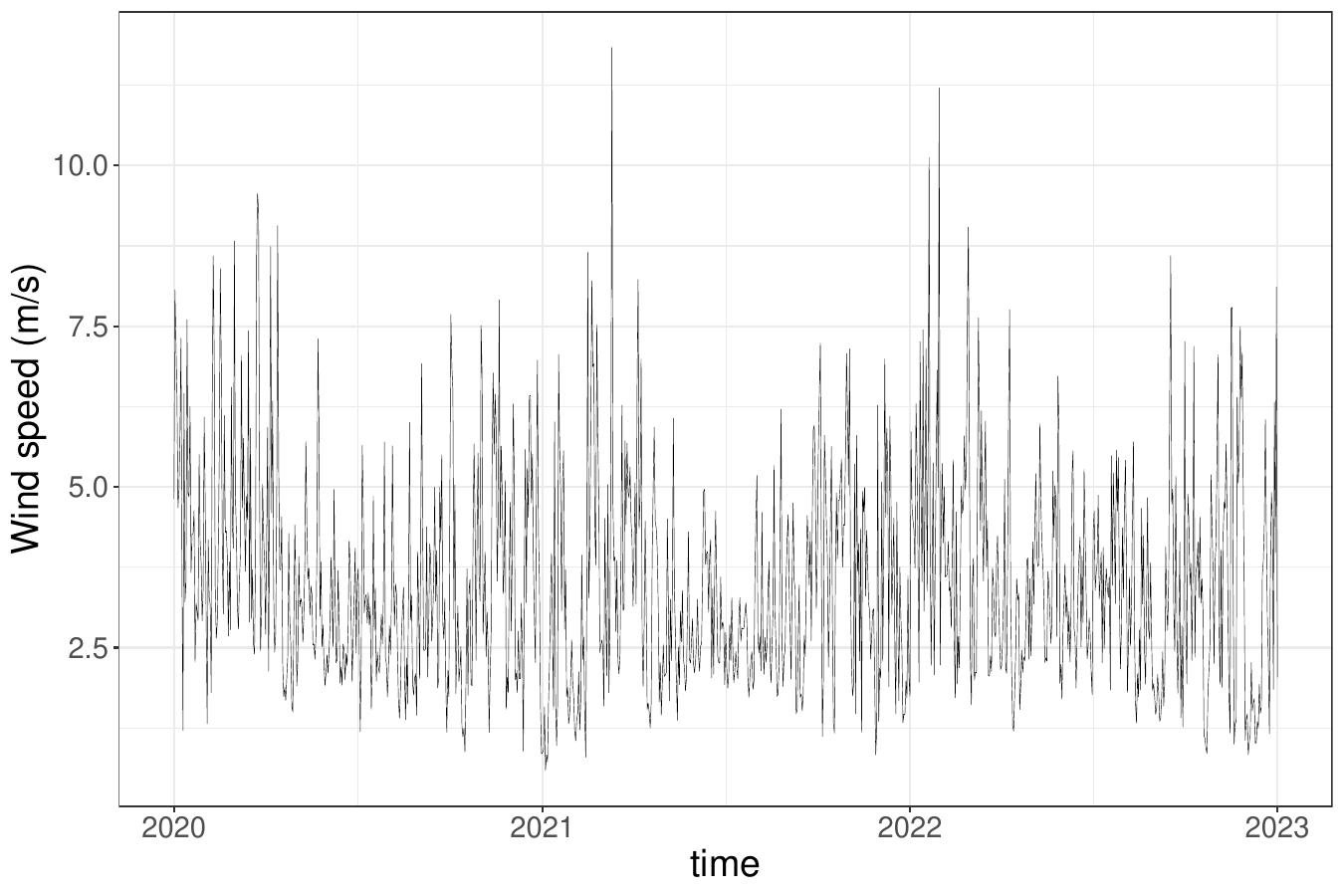}
        \end{subfigure}
        \begin{subfigure}[b]{.31\textwidth}
            \centering
            \includegraphics[width = .95\textwidth]{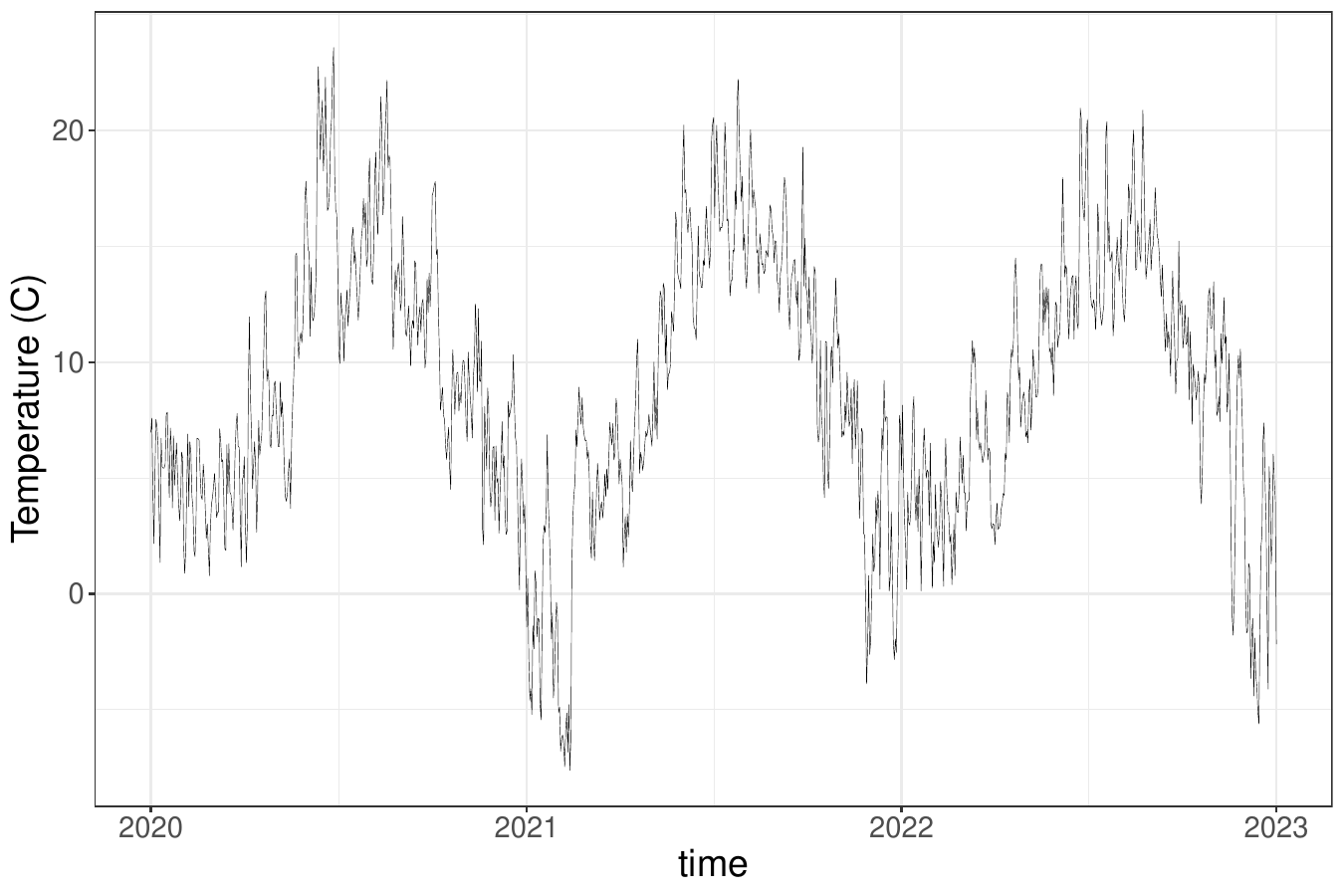}
          \end{subfigure}
          \caption{Observed time series of the available covariates for the whole study period: (a) total precipitation (millimeters), (b) wind speed (m/s) and (c) temperature (C°).}
          \label{fig:obscovariate}
        \end{figure}

To enrich the analysis, these pollution data are merged with meteorological data that are collected hourly from the nearest high-quality weather station located at Florida, approximately one kilometer from Danmarksplass. These meteorological variables have been aggregated on the daily scale to match the pollutant resolutions and can be used as covariates for modeling the pollutant concentrations dynamics.
In particular, we consider the daily total precipitation (in millimeters), daily average wind speed (in meters per second), and daily average temperature (in degrees Celsius), see Figure \ref{fig:obscovariate}. 
Bergen's geographic location -- facing the North Atlantic Ocean and situated along the Byfjorden -- gives rise to an oceanic climate characterized by mild summers, cool and wet winters, and generally high humidity. The proximity to the ocean exerts a thermal buffering effect, moderating seasonal temperature variations: average temperatures hover around $9^\circ$C, with winters being relatively mild and summers cooler than those experienced in more inland regions. Therefore, we expect the daily average temperature to be a reasonable proxy for the seasonal patterns visible in the average level of pollutant concentrations'. Furthermore, the city’s coastal position allows moist Atlantic air to funnel inland via the fjord, resulting in frequent precipitation and moderate wind conditions year-round. During the observation period, the longest recorded spell without precipitation lasted 11 consecutive days, although, on average, rainfall occurred at intervals no longer than three days. As rain and wind are considered two of the major drivers in abruptly reducing the concentration of air pollutants \citep{ouyang2015washing,zhang2018influences}, we expect to see similar patterns in the transitions from high-pollution states to lower ones.
Finally, we also enrich the available data by building up a weekend effect covariate from the recording date. This effect might be useful to adjust for the inherent weekly seasonality of pollution data related to the typical patterns of human activities \citep{TAVELLA2023101662}.




\section{Methods}
\label{sec:meth}
Multivariate time series data can be expressed in the form of a $T \times p$ array of observations, say $\bm{Y}=\lsq\by_1^\top,\dots,\by_T^\top\rsq$ with $\by_t=\lsq y_{tj}\rsq_{j=1}^p$, where each single entry $y_{tj}$ denotes the value of the $j$th outcome at time $t$. These observations are viewed as a realization of a stochastic process $\lcur \bY_t\,:\, t=1,\dots,T \rcur$, whose joint distribution is more tractable when conditioned on hidden states. We model this unobserved heterogeneity through a time dependent hidden process $\bu=\lcur\bu_t:\, t= 1, \ldots, T\rcur$, where $\bu_t=\lsq u_{t1}, \ldots, u_{tK}\rsq$ is a multinomial random variable with one trial and $K$ states. The process is said to be in state $k$ at time $t$ if $u_{tk}=1$. 
The law of $\bm{Y}$ can then be specified by a hierarchical model that combines a parametric conditional distribution of the observations given the hidden states $f(\bm{Y} \mid \bm{u};\, \bm{\theta})$ (the observation process), with a parametric model for the latent chain $p(\bm{u}; \bm{\eta})$ (the hidden process). The corresponding marginal distribution of the observed data can be obtained through marginalization of the hidden process as:
\begin{equation} 
\label{eq:hierachical_model}
    f(\bm{Y};\, \bm{\theta}, \bm{\eta})=\sum_{\bm{u}}f(\bm{Y} \mid \bm{u};\, \bm{\theta})\cdot p(\bm{u};\, \bm{\eta}).
\end{equation}


\subsection{The observation process}
\label{sec:meth_observation process}
Hierarchical time-series models involving hidden states are often simplified by assuming the conditional independence of each observation vector $\bm{y}_t$ given the corresponding hidden state $\bm{u}_t$. Under this assumption, the conditional distribution of the full data conveniently factorizes as $f(\bm{Y} \mid \bm{u})=\prod_{t=1}^{T}\prod_{k=1}^{K}f(\bm{y}_t;\bm{\theta}_k)^{u_{tk}}$, where $\btheta_k$ denotes the vector of parameters associated with state $k$.
This formulation implies that each multivariate observation is independent of past and future observations, conditional on the hidden state at time $t$. Consequently, the observed process is fully characterized by specifying a model for the conditional distribution of $\bm{y}_t$ given $\bm{u}_t$. A popular choice is the multivariate Gaussian distribution $\bm{y}_t \mid u_{tk} = 1\, \sim\, \mathcal{N}\lrnd\boldsymbol{\mu}_k,\, \boldsymbol{\Sigma}_k\rrnd,\, t=1,\dots,T$, where \(\boldsymbol{\mu}_k\) and \(\boldsymbol{\Sigma}_k\) denote the state-dependent mean vector and covariance matrix, respectively. This choice is able to capture the within-state dependence structure among the $p$ components of $\bm{y}_t$. 

However, in many real-world applications -- particularly in environmental contexts -- the assumption that the state-specific mean vector $\boldsymbol{\mu}_k$ remains constant over time can be overly restrictive. Empirical evidence often indicates that multivariate time series exhibit autoregressive dependencies, where current outcomes are influenced by past observations. As a result, assuming conditional independence of observations given only the hidden state may be overly simplistic. To accommodate this additional temporal structure, we relax the conditional independence assumption and instead assume that each observation $\bm{y}_t$ is conditionally independent of the rest of the process given both the hidden state $\bm{u}_t$ and a finite history of past observations up to lag $H$. Under this assumption, the factorization of the observed process distribution becomes: 
\begin{equation*}
\label{eq:observation_process}
    f(\bm{Y} \mid \bm{u})=\prod_{t=1}^{T}\prod_{k=1}^{K}f_k\lrnd\bm{y}_t \mid \bm{y}_{t-1}, \ldots \bm{y}_{t-H};\, \bm{\theta}_k\rrnd^{u_{tk}}.
\end{equation*}
More specifically, we model the conditional distributions using a state-dependent vector autoregressive model of order $H$ (VAR($H$)):
\begin{equation}
\label{eq:cond_observation_process}
    f_k\lrnd\bm{y}_t \mid \bm{y}_{t-1}, \ldots \bm{y}_{t-H}; \bm{\theta}_k\rrnd\,=\, \mathcal{N}\lrnd \bb_k+ \sum_{h=1}^{H} \bm{A}_{hk} \bm{y}_{t-h},\, \bm{\Sigma}_k\rrnd,
\end{equation}
where $\bb_k$ is a $p \times 1$ baseline mean vector and $\bm{A}_{hk}, h = 1, \dots, H$ are $p \times p$ autoregressive coefficient matrices that capture temporal dependence across components and lags within each state.

When exogenous covariates are available in the form of a $T \times J$ matrix $\bm{X} = \lsq \bx_1^\top, \dots, \bx_T^\top \rsq$, we incorporate their influence into the baseline level through a linear regression term $\bb_k\lrnd\bx_t\rrnd=\bb_{k0}+\bm{B}_k\bx_t^{\top}$,
where $\bm{B}_k$ is a $p\times J$ matrix of regression coefficients. In our case study, only meteorological factors have been included, but further applications might also include either policy interventions or demographic information.

Under this framework, the conditional mean of the observation process is dynamically updated based on both historical observations and contemporaneous exogenous variables. We notice that the VAR process within each state $k$ is stable (i.e. stationarity is preserved) whenever the eigenvalues of the corresponding companion matrix
lie strictly within the unit circle. This condition guarantees that if the process were to remain in state $k$ indefinitely, it would exhibit stationary behavior rather than diverging. In particular, if the VAR process of all states $k=1,\dots,K$ is stable then the resulting \textit{mixture of VAR} processes $\lcur \bY_t\,:\, t=1,\dots,T\rcur$ also inherits stability properties \citep{fong2007mixture}.

\subsection{The hidden process}
\label{sec:meth_hidden process}

We assume the hidden process to be a the realization of a $K$-states multinomial chain, which is a discrete-time stochastic process $\bm{u}=\lcur\bm{u}_t\,:\, t=1, \ldots, T\rcur$. Its joint distribution can be expressed through the following conditional factorization:
\begin{equation}
\label{eq:latent_process}
    p\lrnd \bm{u}\rrnd = p\lrnd \bm{u}_1\rrnd\cdot \prod_{t=2}^{T} p\lrnd\bm{u}_{t} \mid \bm{u}_{t-1}, \ldots, \bm{u}_{1}\rrnd,
\end{equation}
where $p(\bm{u}_1)=\prod_{k=1}^K\pi_k^{u_{1k}}$ is the initial distribution at time $t=1$ and $p\lrnd\bm{u}_t \mid \bm{u}_{t-1}, \ldots, \bm{u}_{1}\rrnd$ denotes the conditional distribution at times $t=2,\dots,T$ given the full history of the process. 
The time-dependence properties of the chain can be characterized by specific assumptions on the conditional distributions building up Equation \eqref{eq:latent_process}.
In a \textit{semi-Markov chain}, the conditional distributions are assumed to be conditionally independent on the full history of the process given the time of the last transition. This is conveniently summarized by the state of the chain at time $t-1$, say $k$, and the current dwell time $d_{t-1}$, which is the time spent in $k$ since the last transition.
This behavior is fully captured by the transition probability from any state $k=1,\dots,K$ to state $j=1,\dots,K$ after having sojourned $d$ times in state $k$, which we denote as $\gamma_{kj}(d)=P\lrnd u_{tj}=1\given u_{t-1, k}=1, d_{t-1}=d\rrnd$. Then, $\bm{u}$ is a semi-Markov chain if the conditional distributions can be expressed as:
\begin{equation} 
\label{eq:semi-markov chain}
p\lrnd\bm{u}_t \mid \bm{u}_{t-1}, \ldots, \bm{u}_{1}\rrnd=\prod_{k=1}^{K}\prod_{j=1}^{K}\gamma_{kj}\lrnd d_{t-1}\rrnd^{u_{tkj}}, \qquad t=2,3, \ldots, T,
\end{equation}
where $u_{tkj}=u_{t-1,k}\cdot u_{tj}$.
The specification of the statistical model for a (homogeneous) semi-Markov chain is traditionally completed by assuming a parametric model for the sojourn distribution, say $p_k\lrnd d\rrnd$, and the conditional transition probability $K\times K$ matrix $\Omega=\lsq\omega_{kj}\rsq$, where $\omega_{kj}=P\lrnd u_{tj}=1\given u_{tk}=0, u_{t-1,k}=1\rrnd$ . 
Popular choices for the sojourn distribution are the shifted Poisson, the shifted negative binomial, and the logarithmic \citep[see, e.g.,][]{mhsmm_package}, but any distribution on the positive integers is admissible. The extension to inhomogeneous semi-Markov chains is possible by assuming that the dwell-time distribution depends on a row-profile $\bm{z}^{\top}$ of covariates, say $p_k(d; \bm{z}^{\top})$. However, this formulation is restricted to having covariates influence the dwell-time distribution at the start of a stay, which is then fixed to generate the length of the stay \citep{ricciotti2025zero}. This is an unpleasant restriction in studies where the stay of the system in a specific latent regime is instantaneously influenced by time-varying weather conditions, such as in the pollution study that motivated this work. 

In order to obtain a more convenient parametric specification of the transition probabilities in Equation \eqref{eq:semi-markov chain}, we follow an alternative approach recently suggested by \citet{lagona2025nonhomogeneous}, who induce a model on $p_k(d)$ by by means of the probability of leaving state $k$ after a certain dwell time $d$:
\begin{equation*}
    \label{eq:hazard}
    q_k(d)=P\lrnd u_{tk}=0\given u_{t-1,k}=1, d_{t-1}=d\rrnd, \qquad k=1, \ldots, K.
\end{equation*}
Under this setting, the transition probabilities are obtained as:
\begin{equation}
\gamma_{kj}(d_{t-1})=\begin{cases}
    1-q_{k}(d_{t-1}) & j=k\\
    q_{k}(d_{t-1})\cdot \omega_{kj} & j\neq k 
\end{cases} \qquad k,j=1,\dots,K.
\label{eq:proposal}
\end{equation}
In \eqref{eq:proposal}, $q_k(d)$ plays the role of a discrete hazard function as it indicates the probability of switching state given that the chain has ``survived'' $d$ times in a certain state. Note that the two approaches are essentially equivalent, as there is a one-to-one correspondence between hazards and probability distributions of sojourn times. The former completely specifies the latter as:
\begin{equation}
\label{eq:dwell_distribution}
p_k(d)=q_k(d)\prod_{\delta=1}^{d-1}(1-q_k(\delta)),\quad k=1,\dots,K.
\end{equation}
If the hazard is constant, say $q_k(d)=q_k$, then \eqref{eq:dwell_distribution} reduces to a geometric distribution. If this holds for any state $k=1, \ldots, K$, then $\gamma_{kj}(d)=\gamma_{kj}$ do not depend on the dwell time and $\bm{u}$ reduces to a Markov chain with transition probabilities $\gamma_{kj}$. Otherwise, the Markovianity is lost as the transition probabilities are endogenously updated by the dwell time of the current state. 

Drawing from the survival analysis literature, the hazards can be expressed in a generalized linear model fashion as $g\left(q_{k}(d_{t-1})\right)=\beta_{0 k}+\beta_{1 k}d_{t-1}$, where $g$ is a suitable link function (e.g. the complementary log-log) and the $\beta$s are regression coefficients. This expression yields two main advantages. 
First, the constant hazard model is included as a particular case (when $\beta_{1k}=0$), facilitating tests against the null hypothesis of a geometric dwell time. These tests are more complicated under the traditional approach, because popular sojourn time distributions do not typically include the geometric distribution as a particular case. 
Second, it can be extended to include a profile of time-varying covariates $\bm{z}_t$, not necessarily equal to the profile $\bm{x}_t$ that have been exploited at the observation level, namely 
\begin{equation}
\label{eq:glm_covariates}
g\left(q_{k}(d_{t-1})\right)=\beta_{0 k}+\beta_{1 k}d_{t-1}+\bm{z}_t^{\top}\bm{\beta}_{2k}, \qquad k=1, \ldots, K.
\end{equation}
Under \eqref{eq:glm_covariates}, the hazard is endogenously updated by the current dwell time and, exogenously, by a profile of time-varying covariates. By plugging this dynamic hazard regression into Equation \eqref{eq:dwell_distribution}, we obtain a sojourn time distribution that can be modulated by covariates at every time, and not just at baseline, therefore allowing time-varying covariates to influence the sojourn distribution. Under this setting, the joint distribution of $\bm{u}$ is fully specified by a parametric model $p(\bm{u} \mid \bm{\eta})$ that depends on the parameter vector $\bm{\eta}=(\bm{\beta},\bm{\omega})$, where the regression coefficients $\beta$ of the hazard regressions \eqref{eq:glm_covariates} drive the sojourn distribution of each state, whereas the conditional switching probabilities $\omega$ steer the chain to a specific state at each transition.

\section{Likelihood-based inference}
\label{sec:lik}
For a given $K$, the joint distribution of the proposed hierarchical model is known up to the parameters $\bm{\theta}$ and $\bm{\eta}$, where $\bm{\theta}=\lcur\btheta_k\,:\,k=1,\dots,K\rcur$
are the observation process parameters, and $\bm{\eta}=\lcur\bm{\pi},\bm{\beta},\bm{\Omega}\rcur$ are the hidden process ones. Parameter estimation given the observed data for a fixed number of hidden states $K$ and maximum auto-regressive lag $H$ can be pursued by optimizing the marginal likelihood corresponding to Equation \eqref{eq:hierachical_model}:
\begin{equation*}
\label{eq:M_likelihood}
L(\bm{\theta},\bm{\eta})=
\sum_{\bm{u}}\prod_{t=1}^{T}\prod_{k=1}^{K}f_k(\bm{y}_t \mid \bm{y}_{t-1}, \ldots \bm{y}_{t-H}; \bm{\theta}_k)^{u_{tk}}\cdot p(\bm{u}_1;\bm{\pi})\prod_{t=2}^{T}\prod_{k=1}^{K}\prod_{j=1}^{K}\gamma_{kj}(d_{t-1};\omega_{kj}, \bm{\beta}_k)^{u_{tkj}}.
\end{equation*}
However, its direct maximization is hindered by the overwhelming summation over all the possible paths of the latent semi-Markov chain. This difficulty can be overcome through the approximation of the hidden semi-Markov chain as a hidden-Markov one with $K\times m$ number of states, where $m$ is the maximum dwell-time for which the parametric assumption of Equation \eqref{eq:glm_covariates} holds. Dwell times larger than $m$ are still allowed, but their distribution is approximated by a geometric tail \citep{zucchini2009hidden, lagona2025nonhomogeneous}. 
The HMM formulation gives way to the implementation of a workable EM algorithm where maximization of a weighted complete-data log-likelihood (M step) is iteratively alternated with weights updating (E step), until convergence. Convergence is assessed through log-likelihood- and/or parameter-based stopping criteria. 

Specifically, at each iteration $\ell$, the E-step evaluates the univariate posterior probabilities $\hat{\pi}_{tkd}^{(\ell)} = P(u_{tkd}=1\mid \bm{y},\bm{x},\bm{z}, \bm{\theta}^{(\ell-1)}, \bm{\eta}^{(\ell-1)})$ and the bivariate posterior probabilities $\hat{\pi}_{tkdjd^{\prime}}^{(\ell)}
=P(u_{t-1,kd}=1, u_{tjd^{\prime}}=1 \mid \bm{y},\bm{x},\bm{z}, \hat{\bm{\theta}}^{(\ell-1)}, \hat{\bm{\eta}}^{(\ell-1)})$ through the \textit{forward-backward} recursions of the Baum-Welch algorithm (see \cite{Cappe_etal2005} for an excellent review).
Such weights are exploited by the subsequent M-step that updates the parameters by maximizing the weighted log-likelihood function:
\begin{equation*}  \label{eq:exp.complete.loglikelihood}
    \begin{aligned}
        Q(\bm{\theta}, \bm{\pi},\bm{\omega},\bm{\beta}) =& \underbrace{\sum_{k=1}^{K}\hat{\pi}_{1k}^{(\ell)}\log \pi_k}_{Q(\bm{\pi})}+\underbrace{\sum_{t=2}^{T}\sum_{k=1}^{K}\sum_{j=1}^{K}\sum_{d,d^{\prime}=1}^{m} \hat{\pi}_{tkdjd^{\prime}}^{(\ell)}\log \gamma_{kj}(d,\bm{z}_t; \bOm, \bm{\beta})}_{Q(\bOm, \bm{\beta})}\\
        &+\,\underbrace{\sum_{t=1}^{T}\sum_{k=1}^{K}\hat{\pi}_{tk}^{(\ell)}\log f_k(\bm{y}_t\mid \bm{y}_{t-1}, \ldots \bm{y}_{t-H};\bm{\theta}_{k})}_{Q(\bm{\theta})},
    \end{aligned}
\end{equation*}
where $Q(\bOm, \bm{\beta}) = Q(\bOm) + Q(\bm{\beta})$ with $Q(\bOm)=\sum_{t=2}^{T}\sum_{k=1}^{K}\sum_{j \neq k}\sum_{d,d^{\prime}=1}^{m}\hat{\pi}_{tkdjd^{\prime}}\log \omega_{kj}$ and:
\begin{equation*}
\begin{aligned}
Q(\bm{\beta})=\sum_{t=2}^{T}\sum_{k=1}^{K}\sum_{h \neq k}\sum_{d,d^{\prime}=1}^{m} \hat{\pi}_{tkdjd^{\prime}}\log q_k(d,\bm{z}_t;\bm{\beta}) +\sum_{t=2}^{T}\sum_{k=1}^{K}\sum_{d,d^{\prime}=1}^{m} \hat{\pi}_{tkdkd^{\prime}}\log (1-q_k(d, \bm{z}_t;\bm{\beta})).
\end{aligned}
\end{equation*}

The weighted log-likelihood function is the sum of functions that depend on independent sets of parameters, which can be maximized separately. Functions $Q(\bm{\pi})$ and $Q(\bOm)$ are weighted multinomial log-likelihoods for which the points of maximum are available in closed form as:
\begin{equation*}
    \hat{\pi}^{(\ell)}_k=\hat{\pi}^{(\ell)}_{1k},\quad \hat{\omega}_{kj}^{(\ell)}=\frac{\sum_{t=2}^{T}\sum_{j \neq k}\sum_{d,d^{\prime}=1}^{m}\hat{\pi}^{(\ell)}_{tkdjd^{\prime}}}{\sum_{t=2}^{T}\sum_{k=1}^{K}\sum_{j \neq k}\sum_{d,d^{\prime}=1}^{m}\hat{\pi}^{(\ell)}_{tkdjd^{\prime}}}.
\end{equation*}
$Q(\bm{\beta})$ is a Binomial log-likelihood that can be maximized by conventional iteratively reweighted least-squares routines, leading to $\hat{\bm{\beta}}_k^{(\ell)}$.   

Function $Q(\bm{\theta})$ is the weighted log-likelihood of a VAR model and can be maximized via the SUR (Seemingly Unrelated Regression) method. For each $j$ and $k$, the optimal estimate of the VAR parameters 
 $\lcur\hat{\ba}_{jk}, \hat{b}_{0jk}, \hat{\bb}_{jk}\rcur^{(\ell)}$
are the minimizers of the weighted sum of squares $\bm{e}_{jk}^{\sf T}W^{(\ell)}_{k}\bm{e}_{jk}$, where $\bm{e}_{jk}=\lsq e_{1jk}, \ldots e_{Tjk}\rsq$ is the residual vector of model \eqref{eq:cond_observation_process}:
\begin{equation}
\label{eq:VAR_residuals}
e_{tjk}=\left(y_{tj}-b_{0jk}-\bm{x}^{\sf T}\bm{b}_{jk}-\sum_{h=1}^{H}\sum_{j=1}^{J}a_{jkh}y_{t-h,j}\right)
\end{equation} 
and $W^{(\ell)}_k$ is the $T \times T$ diagonal matrix of the weights $\hat{\pi}^{(\ell)}_{tk}=\sum_d\hat{\pi}^{(\ell)}_{tkd}$. The corresponding estimate of $\hat{\Sigma}_k^{(\ell)}$ can be obtained by plugging these estimates into the residuals \eqref{eq:VAR_residuals} and computing each entry as:
\begin{equation*}
\hat{\sigma}_{ijk}=\frac{\sum_{t=1}^{T}\hat{\pi}_{tk}\cdot e_{tik}e_{tjk}}{\sum_{t=1}^{T}\hat{\pi}_{tk}}.
\end{equation*}
However, such SUR approach involves a large number of parameters and requires the selection of the lags to be included in the VAR specification. These are generally unknown and would require post-hoc model selection scheme. 
Therefore, to enhance interpretability and avoid the arbitrary choice of the lags, we regularize the least squares by a LASSO penalty   \citep{basu2015highdimensionalvar, Tan21102021}, updating the VAR parameters as the solution of the penalized least squares problem:
\begin{equation}
\mathrm{min} \left\{ 
\bm{e}_{jk}^{\sf T}W^{(\ell)}_{k}\bm{e}_{jk}
+ \lambda_{k}\cdot \| \ba_{jk} + \bb_{jk} \|_1 \right\},
\end{equation}
using a battery of penalty coefficients $\lambda_k$ that operate on the state-specific regression coefficients of the VAR. 
For a sufficiently large value of the maximum lag $H$, this LASSO regularization enforces sparsity and ensures that the irrelevant lagged variables and covariates are automatically excluded from the model, leading to a more parsimonious representation.

 The proposed estimation procedure leads to an optimal solution for a specific number $K$ of hidden states $K$ and a specific value of the  penalty coefficients $\lambda_k, k=1,\dots,K$. Their choice is nontrivial, especially under a mixture setting, where a post-hoc selection over a grid of possible combinations of the two is compared through goodness-of-fit or prediction metric. In our case, we consider and test through a simulation study the use of the Integrated Complete Likelihood (ICL, see \cite{Biernacki2000}) to perform model selection in terms of both these terms. 
However, we must note that in a mixture context the size of the grid of $\lambda_k$s on which the validation process should be performed increases quadratically with the number of states $K$, becoming rapidly overwhelming even for moderate $K$. This issue is particularly pronounced in an HSMM context, where even a single fit might be computationally demanding. 
One could consider adopting one single value $\lambda_k=\lambda$, but it is well known that the optimal penalty value in penalized regression usually depends on sample size \citep{buhlmann2011statistics}.
Therefore, the penalty parameter magnitude should at least account for the effective number of units belonging to each state. In the context of finite mixture of regression, \cite{khalili2007variable} proposes penalties that depend on the size of the regression coefficients and the mixture structure. Similarly, \cite{stadler2013penalized} proposes a penalty that automatically and dynamically adapts to the current state-specific effective sizes within the EM optimization routine. In the latter, the dynamically varying penalty term is specified as:
\begin{equation*}
    \lambda_k= \lambda_0\cdot\sqrt{\sum_{t=1}^T\hat{\pi}_{tk}}, \quad k = 1, \dots, K.
\end{equation*}
$\lambda_0$ is a baseline penalty term susceptible to state-specific penalty adjustments which are proportional to the square root of the \textit{effective sample size} of each state, conforming to the asymptotic theory on the optimal penalty developed in \cite{stadler2010}.

The proposed EM algorithm is stopped when the relative variation in all the parameters between two consecutive iterations is lower than $10^{-4}$. This yields the final parameters' estimates $\lrnd\hat{\btheta},\hat{\boeta}\rrnd$. In such a complex context the derivation of uncertainty of the parameters through the inversion of the Hessian is not practical. Therefore, we quantify uncertainty through the implementation of a parametric bootstrap \cite{efron2000bootstrap}. This is further complicated by the LASSO penalty, for which the bootstrap would return biased confidence intervals of the true parameters \citep{chatterjee2011bootstrapping, li2020debiasing}. Therefore, we first de-bias the final estimates by mean of a \textit{relaxed} LASSO fit (i.e. one more round of the M-step in EM with only the selected coefficient and no penalty), obtaining $\hat{\btheta}^R$. Then, we simulate $B$ datasets from the joint model with parameters $\lrnd\hat{\btheta}^R,\hat{\boeta}\rrnd^R$ and re-fit the whole procedure, model selection included, on each of them. This yields $\lrnd\hat{\btheta}^{b},\hat{\boeta}^{b}\rrnd^B_{b=1}$ that can be used to quantify the overall uncertainty of $\lrnd\hat{\btheta},\hat{\boeta}\rrnd$.

\section{Simulation study}\label{sec:simstudy}

We devise a simulation study to assess and verify the validity of our proposal in three major aspects: (i) evaluate whether the ICL can effectively discriminate the best pair $\lrnd K, \lambda_0\rrnd$; (ii) recovery of the model parameters and selection of the non-zero coefficients in the observation process; (iii) time-series segmentation. 

We simulate $B=200$ replicate datasets with $n = 1100$ observations to match the size of the real data under analysis. The outcome dimension is set to $p = 3$ and the maximum lag order equal to $H=4$, for a varying number of mixture components $K= 2, 3, 4$. 
The autoregressive coefficients have been chosen such that each state corresponds to a stable VAR and the effect at lags going from $1$ to $3$ is decreasing within each outcome and state. The $4$-th lags do not necessarily respect this decreasing behavior as they could represent seasonality patterns. 
The correlation matrices have been randomly generated from an Inverse-Wishart distribution centered around identity matrices of conforming size and $\nu=3$ degrees of freedom. 
The dwell time parameters have been set to values that ensure different sojourn distributions. 
Furthermore, we include the effect of one randomly generated (mean-zero Gaussian) time-varying covariate on the observation and the hidden process.
This setup simulates realistic scenarios with partially overlapped mixture components exhibiting a wide range of correlations and different behaviors of the latent components. In all scenarios, we fit the models for varying number of hidden states $\tilde{K}= 2, 3, 4$ and a sequence of values starting from $0$ and followed by 20 equally spaced values of $\log(\lambda_0)$ between $10^{-4}$ and $0.05$.

We first verify wether the $\tilde{K}^*$ identified as optimal by the ICL matches the true size across all scenarios. 
Table \ref{tab:sim_modelsel} reports the average ICL for each combination of simulated and estimated model. 
\begin{table}
    \centering
    \caption{Average ICL scores for model selection in the simulation study.}
    \label{tab:sim_modelsel}
    \begin{tabular}{cc|rrr}
    \toprule
    &     & \multicolumn{3}{c}{Estimated} \\
    & $K$ & \multicolumn{1}{c}{2} & \multicolumn{1}{c}{3} & \multicolumn{1}{c}{4} \\
    \midrule 
    \multirow{3}{*}{\rotatebox[origin=c]{90}{Simulated}} & 
         2  & \textbf{10854.56} & 11084.04 & 11342.60  \\
      &  3  & 11974.74  & \textbf{11432.90} & 11675.66 \\
      &  4  & 11475.18  & 11314.95 & \textbf{10858.25}\\
         \bottomrule
    \end{tabular}
\end{table}
Results show that the average ICL is lowest for the model with the \textit{true} number of components, highlighting the ability of the ICL to select the correct model specification also under a penalized regression framework. In addition, comparing the ICL of each single replica, we report that the correct model (i.e. $\tilde{K}^*=K$) is selected 100\% of the time across all $K$s.
Therefore, we can focus on the performances of the well-specified models only (i.e. $K=\tilde{K}^*$) and, in particular, we report in the main text the results for the $K=2$ scenario for reason of space. The results for $K=3,4$ are briefly commented here but detailed in the Supplementary Material.

\begin{figure}[t]
    \centering
    \includegraphics[width=0.8\textwidth]{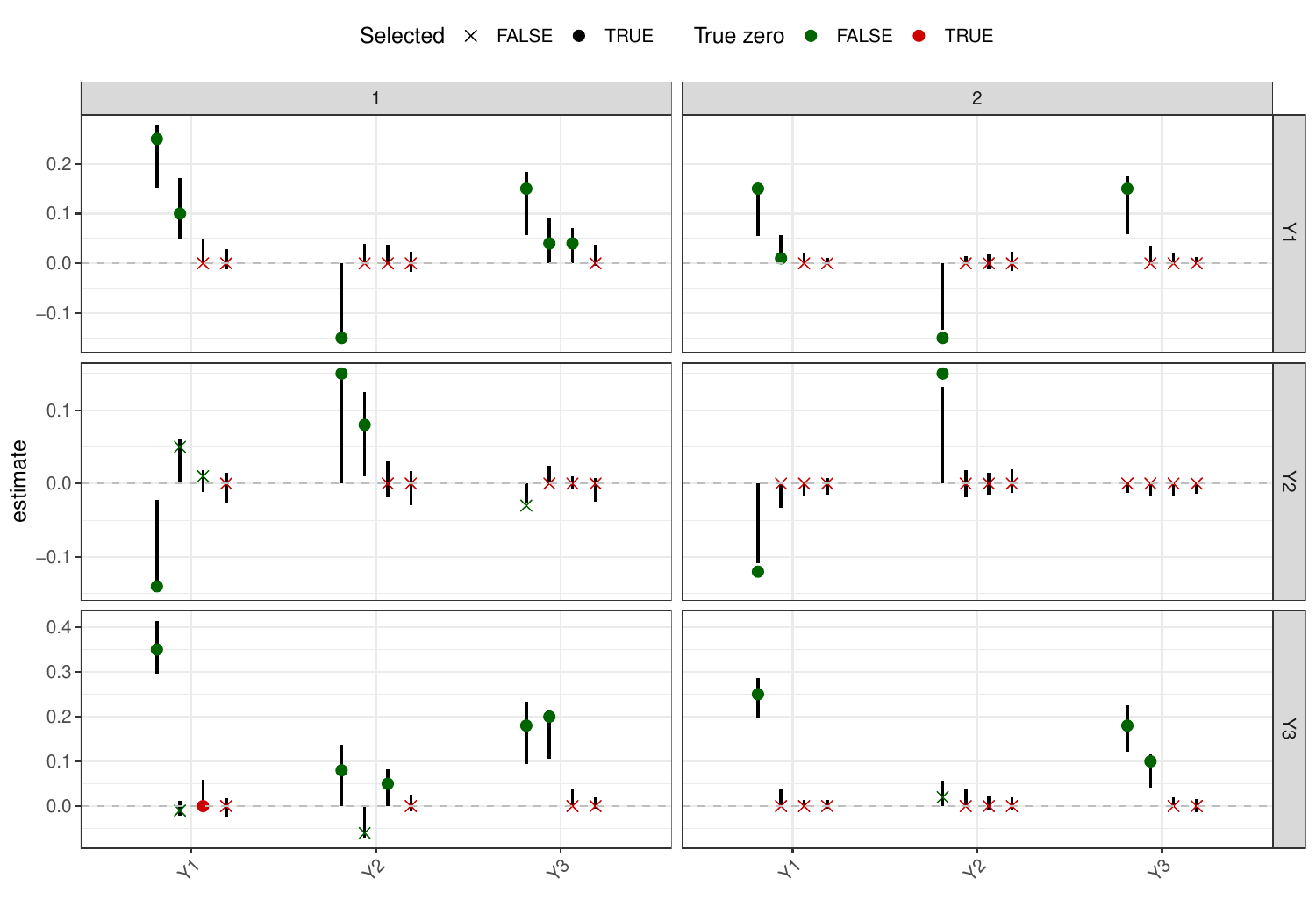}
    \caption{Recovery of the auto-regressive coefficients for $K = 2$. True values are: green if different from $0$, red if not; marked as a cross if shrunk to $0$ more than $50\%$ of the times and as a dot otherwise. The black line represents the central $95\%$ of the estimates distribution.}
    \label{fig:sim_VAR2}
\end{figure}

To evaluate the recovery of the regression and auto-regressive coefficients of the observation process, other than the optimal selection of $\lambda_0^*$, we compare the bootstrapped distribution of the estimates with their true values. Table \ref{tab:simresbetaVAR2} reports some relevant statistics on the regression coefficients, while the performances on the auto-regressive coefficients are summarized in Figure \ref{fig:sim_VAR2}. 
\begin{table}
    \centering
        \caption{Recovery of the regression coefficient for $K=\tilde{K}^*=2$: True value, Mean of the estimates, $95\%$ central interval, Root Mean Squared Error (RMSE), proportion of selection.}
    \adjustbox{max width = .95\textwidth}{%
    \begin{tabular}{c|rrrrr|rrrrr}
    \toprule
      \multirow{2}{*}{$\bB$}  & \multicolumn{5}{c|}{$k=1$} & \multicolumn{5}{c}{$k=2$} \\
       & True & Mean & \multicolumn{1}{c}{CI$_{0.95}$}  & RMSE & \multicolumn{1}{c|}{\%} & True & Mean & \multicolumn{1}{c}{CI$_{0.95}$} &  RMSE & \multicolumn{1}{c}{\%} \\
        \midrule
       $b_{01}$ & 3 & 3.11 & (2.97, 3.26) & 0.137 & -- & -1 & -0.88 & (-1.02, -0.75) & 0.138  & -- \\
       $b_{02}$ & 1.5 & 1.38 & (1.23, 1.57) & 0.145 & -- & -2 & -2.14 & (-2.29, -2.00) & 0.153  & -- \\
       $b_{03}$ & 2 & 2.01 & (1.85, 2.14) & 0.081 & -- & -1.5 & -1.49 & (-1.58, -1.38) & 0.057  & --\\
       $b_{11}$ & 0.5 & 0.457 & (0.383, 0.548) & 0.061 & 100 & -0.2 & -0.158 & (-0.249, -0.076) & 0.060  & 100    \\
       $b_{21}$ & 0.0 & 0.001 & (-0.039, 0.053) & 0.022 & 35 & 0.4 & 0.357 & (0.287, 0.438) & 0.059 & 100    \\
       $b_{31}$ & 0.0 & -0.002 & (-0.062, 0.051) & 0.023  & 41 & -0.1 & -0.057 & (-0.129, 0.000) & 0.058 & 87 \\
    \bottomrule         
    \end{tabular}
    }
    \label{tab:simresbetaVAR2}
\end{table}
The coefficients' selection performs well across all outcomes and states, with a few exceptions that are characterized by selection percentages close to $50\%$. In particular, the estimates of the coefficients that are wrongly selected as non-zero always present near-zero point estimates and zero-overlapping intervals. 
The distribution of the non-zero coefficients is slightly biased toward lower values, which is a common issue in penalized regression, but the true value is always contained in the $95\%$ Central Intervals.
The recovery of the Variance-Covariance matrices is evaluated through the matrix Kullback-Leibler (KL) discrepancy, whose average values are very low and equal to $0.014$ and $0.012$ for $K=1,2$, respectively.
Similar performances are observed for the other $K$s (see Supplementary Material).
Table \ref{tab:simresbeta} shows the estimates of the dwell time coefficients.
\begin{table}
    \caption{Recovery of the dwell-time regression coefficients $K=\tilde{K}^*=2$: True value, Mean of the estimates, $95\%$ Central Interval, Root Mean Squared Error (RMSE).}
    \centering
    \adjustbox{max width=0.95\textwidth}{%
    \begin{tabular}{c|cccc|cccc}
    \toprule
      \multirow{2}{*}{$\boeta$} & \multicolumn{4}{c}{$k=1$} & \multicolumn{4}{c}{$k=2$} \\
       & True & Mean & CI$_{0.95}$ & RMSE & True & Mean & CI$_{0.95}$ & RMSE \\
       \midrule
       $\beta_0$ & -1 & -1.03 & (-1.35, -0.76) & 0.16  & -2 & -2.03 & (-2.38, -1.71) & 0.18 \\
       $\beta_1$ & 0.15 & 0.16 & (0.07, 0.25) & 0.05 & 0.35 & 0.36 & (0.28, 0.46) & 0.05  \\
       $\beta_2$ & -0.50 & -0.51 & (-0.67, -0.38) & 0.07 & 0.50 & 0.50 & (0.36, 0.66) & 0.07 \\
         \bottomrule         
    \end{tabular}
    }
    \label{tab:simresbeta}
\end{table}
The point estimates yield averages that closely align with the true values, which consistently fall within the 95\% Central Intervals. The RMSEs remain low across all cases, demonstrating that our method effectively captures the dynamics of the latent terms. The conditional transition probability matrix $\Omega$ is not estimated for $K = \tilde{K}^* = 2$, as it is deterministically defined as a row-reversed identity matrix $J_K$.

Regarding time-series segmentation, each observation is assigned a cluster label with the Maximum A Posteriori (MAP) rule. The average ARI and Accuracy are notably high (exceeding 87\% and 95\%, respectively) in all scenarios. This underscores the model’s ability to accurately assign each data point to its corresponding cluster, regardless of model complexity. Detailed results are included in the Supplementary Material.

\section{Application}\label{sec:application}

\begin{figure}
    \centering
    \includegraphics[width=0.8\linewidth]{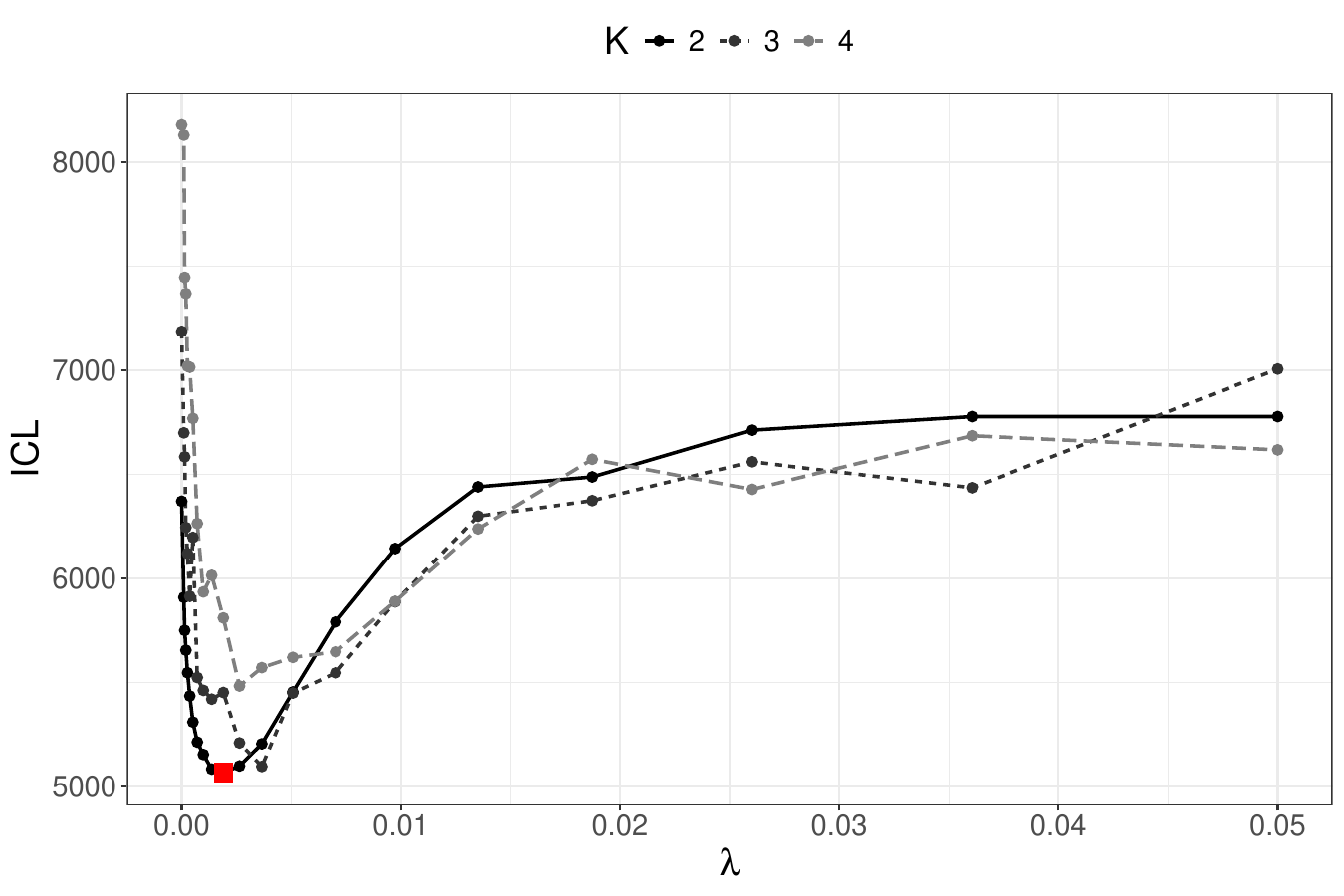}
    \caption{Model selection: cross-validation of $\lambda$ for each $K$.}
    \label{fig:iclsel}
\end{figure}

The model is applied to the data described in Section 2 for different values of $K= 2, 3, 4$ and $\lambda_0= 0, e^{10^{-4}},\dots,e^{0.05}$, where the exponents from $10^{-4}$ and $0.05$ are equally spaced. The order of the VAR is set to $H=7$ across all pollutants and states, so that all reasonable lags up to those responsible for weekly seasonality could be selected. We use the average daily temperature and the weekend effect (namely, weekend = 1 if the day of the week is Saturday or Sunday, weekend = 0 otherwise) as exogenous time-varying covariates for the mean term of the observations process. 
We instead consider the average wind speed and total precipitation as time-varying covariates on the hidden process hazards. These are well-known environmental conditions that affect the chances of reducing pollution levels by either \textit{diluting} their concentrations into larger areas or washing them out, that is, dragging them to the ground. 
As a link function for the hazard $g$, we use the c-loglog, which is the canonical link resulting from the specification of a Gompertz-type hazard in discrete time. 
Uncertainty is quantified through bootstrapping over $B=300$ simulated sets.
The optimal penalty $\lambda^*$ and number of states $K$ is selected via cross-validation according on the ICL. The results are summarized in Figure \ref{fig:iclsel}.
The behavior for each $K$ is convex, as expected, and the best score is achieved for the $K = 2$, at the $11$-th value of $\lambda$. Therefore all the following results will refer to this model's specification.

\begin{figure}[t]
        \centering
        \begin{subfigure}[b]{.3\textwidth}
            \centering
            \includegraphics[width = .99\textwidth]{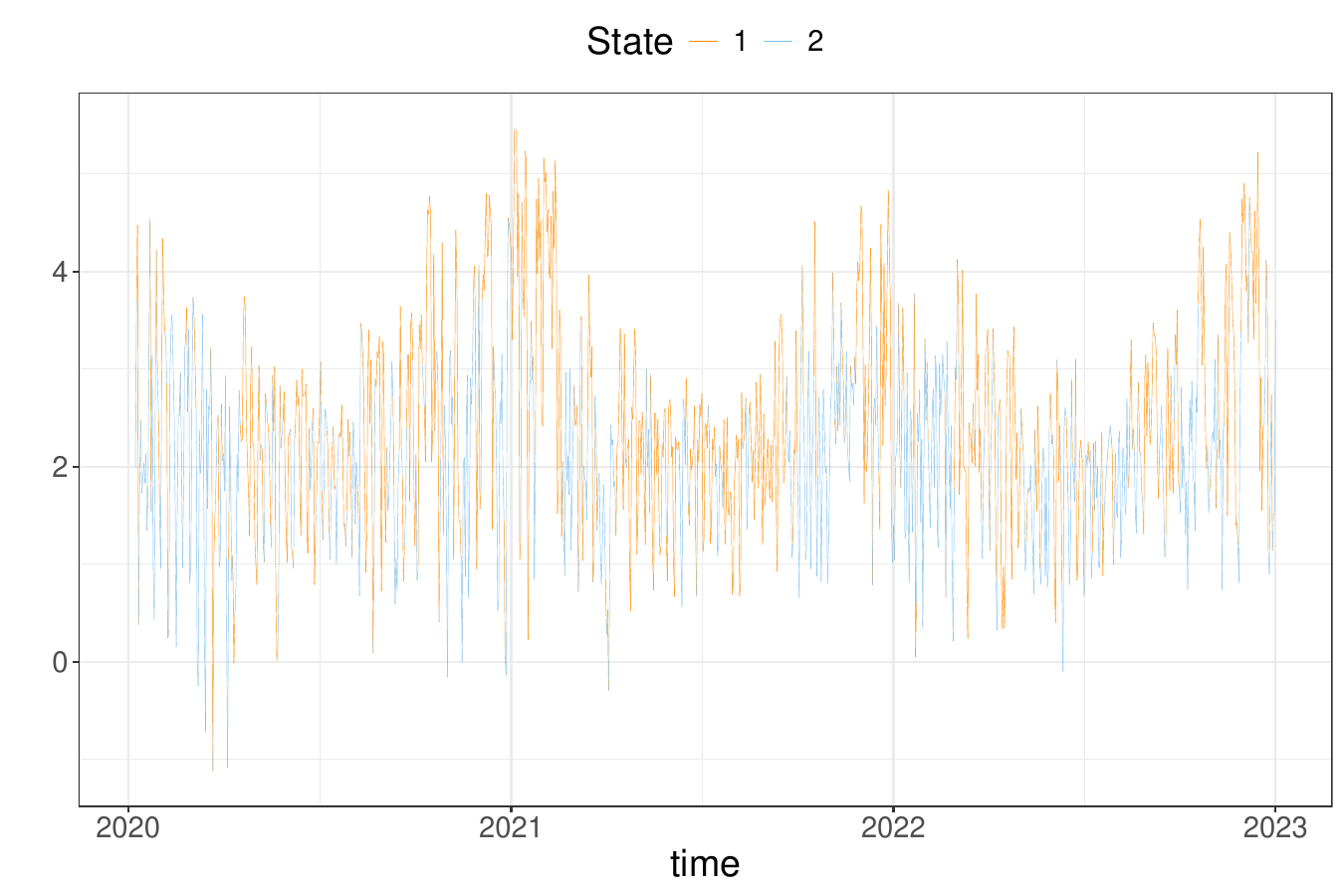}
            \caption*{\scriptsize NO}
        \end{subfigure}
        \begin{subfigure}[b]{.3\textwidth}
            \centering
            \includegraphics[width = .99\textwidth]{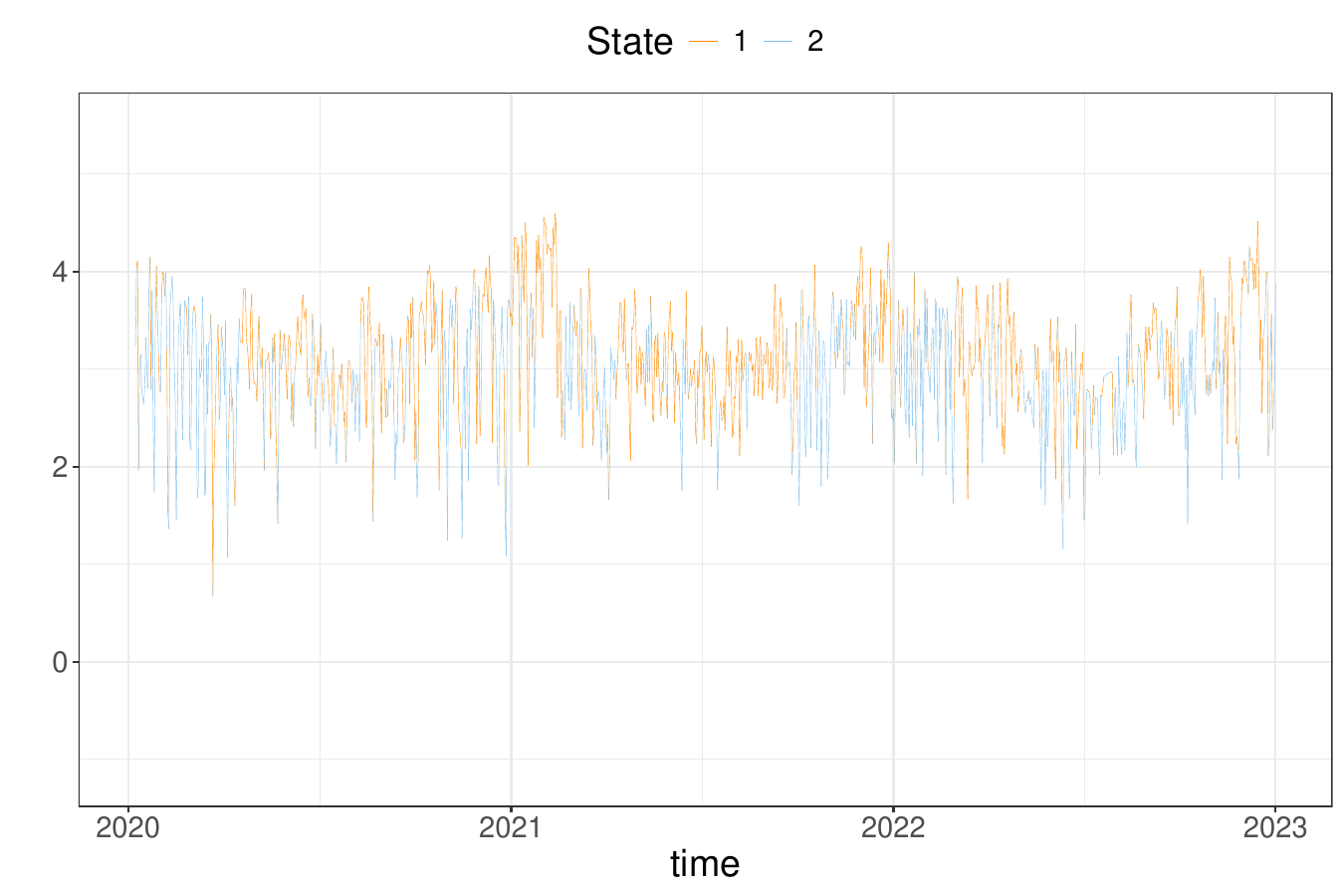}
            \caption*{\scriptsize NO$_2$}
        \end{subfigure}
        \\
        \begin{subfigure}[b]{.3\textwidth}
            \centering
            \includegraphics[width = .99\textwidth]{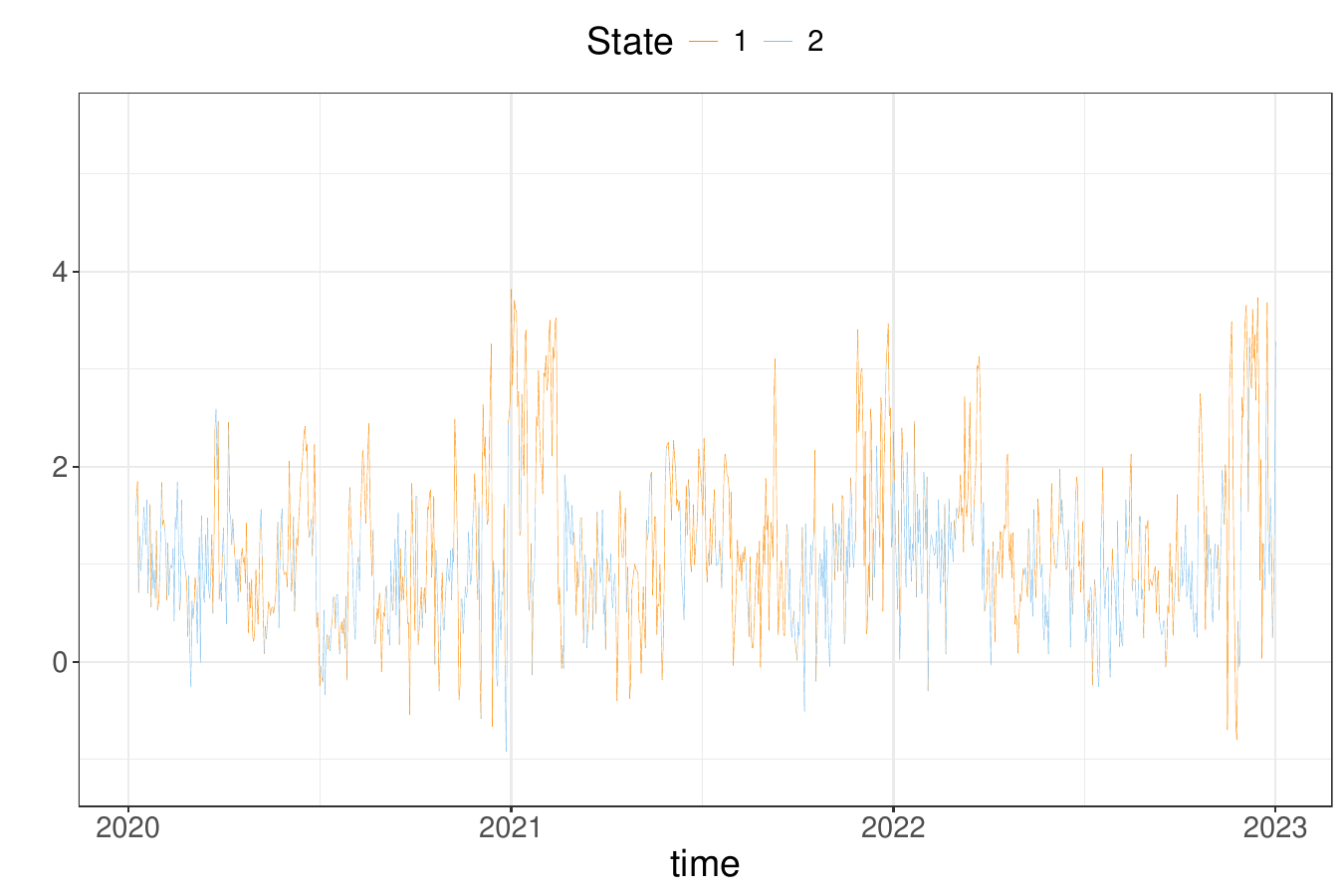}
            \caption*{\scriptsize PM$_1$}
        \end{subfigure}
        \begin{subfigure}[b]{.3\textwidth}
            \centering
            \includegraphics[width = .99\textwidth]{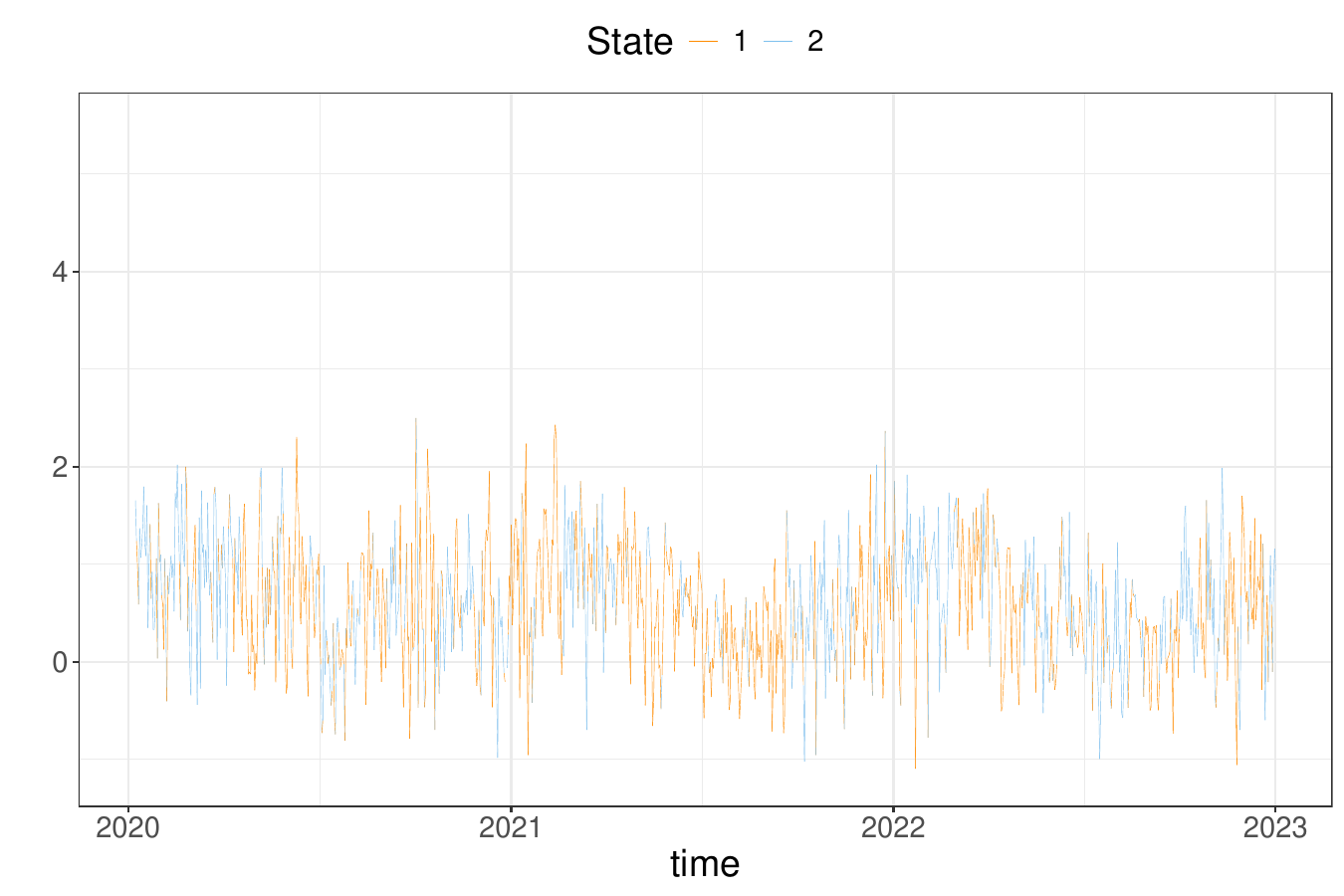}
            \caption*{\scriptsize PM$^*_{2.5}$}
        \end{subfigure}
        \begin{subfigure}[b]{.3\textwidth}
            \centering
            \includegraphics[width = .99\textwidth]{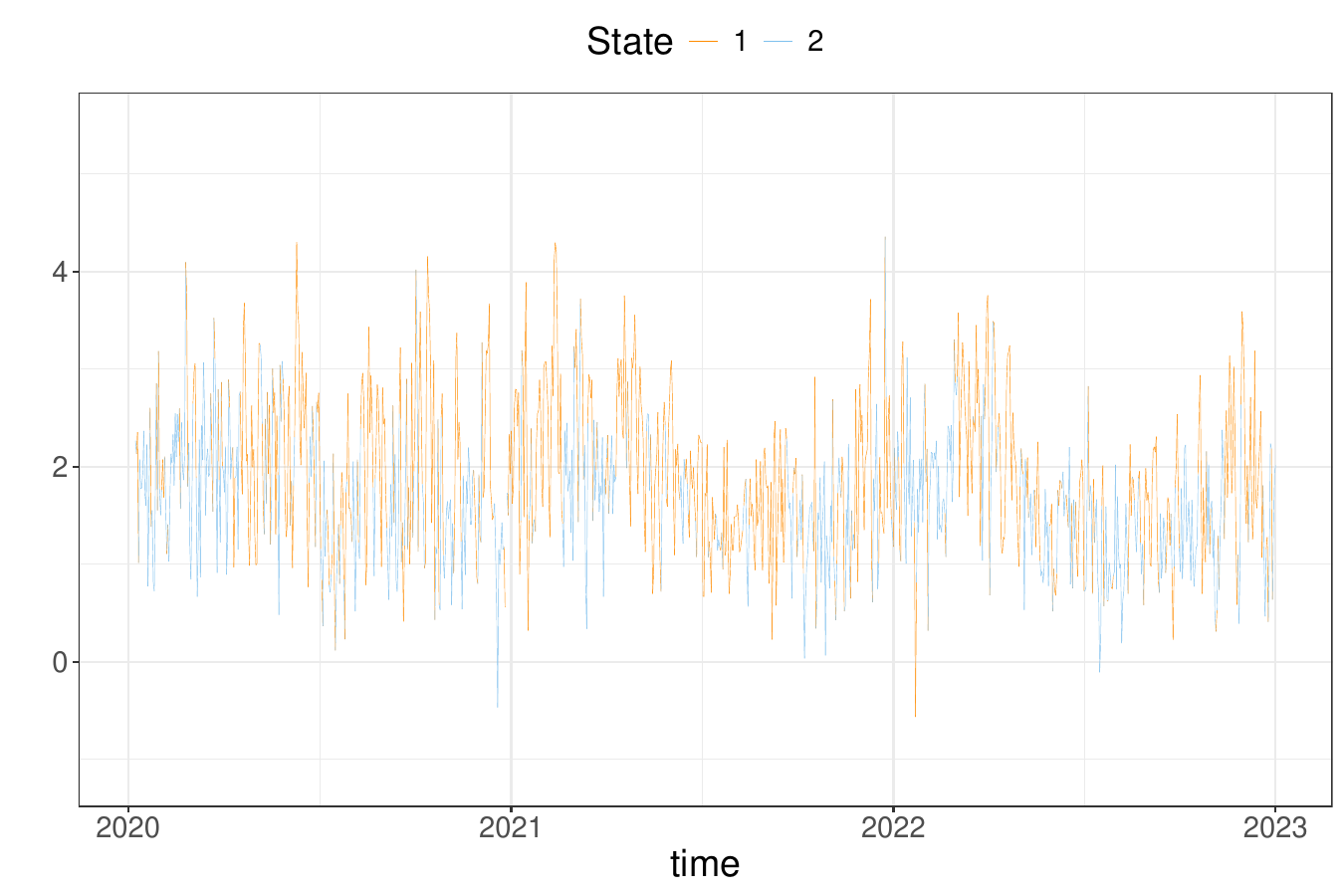}
            \caption*{\scriptsize PM$^*_{10}$}
        \end{subfigure}
        \caption{Estimated time-series segmentation.}
        \label{fig:ts_segment}
    \end{figure}

\begin{figure}[t]
    \centering
    \includegraphics[width=0.85\linewidth]{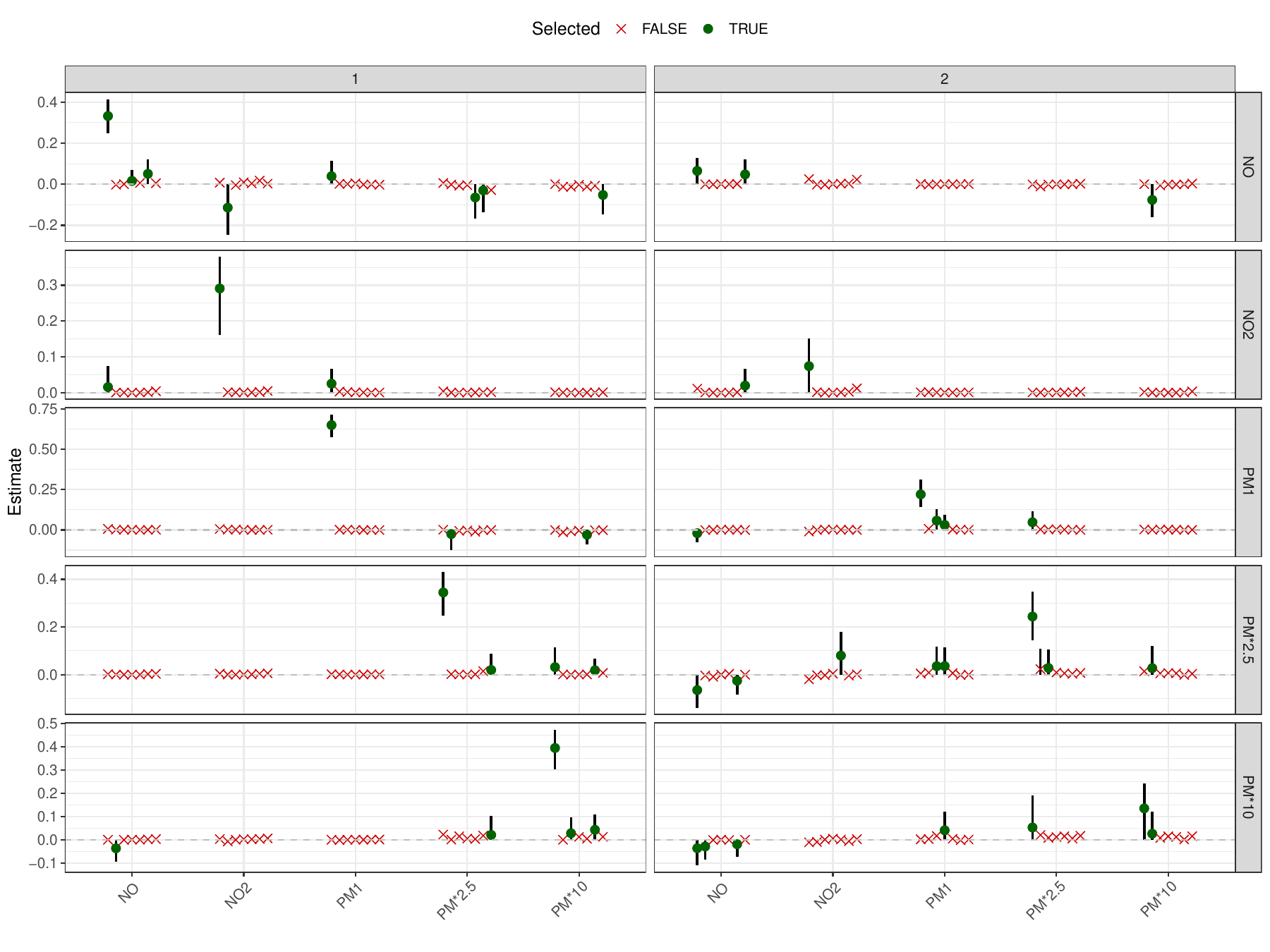}
    \caption{Estimated auto-regressive coefficients. Estimates are denoted with a  green dot if selected for more of $50\%$ of the bootstrap sample and red crosses otherwise. The black line represents the bootstrapped $95\%$ Confidence Intervals.}
    \label{fig:estdvarcoef}
\end{figure}

Figure \ref{fig:ts_segment} presents the estimated time-series segmentation over the entire study period.
The VAR of both states is stable for both states, with state 1 (orange) corresponding to higher pollution levels and state 2 (sky blue) to lower ones. Indeed, the marginal means of the corresponding VAR processes (net of the covariates effects) are $\mu_1=\lsq 3.76, 3.70, 1.80, 1.00, 2.76\rsq$ and $\mu_2=\lsq 2.34, 3.24, 0.84, 0.67, 1.77\rsq$), respectively.
Pollutants are in state 1 for 55\% of the days, most of which occur during spring and summer. State 2 occurs approximately as frequently as state 1, but is predominantly observed in autumn and winter. This is primarily due to the type of heating systems used in residential and commercial buildings, as well as a greater reliance on motorized transport, given the challenges posed by adverse weather conditions for walking or cycling. 
Table \ref{tab:resestcovar} reports the estimated regression coefficients on the observation process.
\begin{table}
    \centering
     \caption{Estimated regression coefficients on the observations process: estimates, $95\%$ confidence intervals, percentage of selection.}
    \label{tab:resestcovar} 
      \adjustbox{max width=\textwidth}{%
    \begin{tabular}{lr|rrrrrrrrrr}
    \toprule
     &  & \multicolumn{2}{c}{NO} & \multicolumn{2}{c}{NO$_2$} & \multicolumn{2}{c}{PM$_{1}$} & \multicolumn{2}{c}{PM$^*_{2.5}$} & \multicolumn{2}{c}{PM$^*_{10}$} \\
      & $k$ & \multicolumn{1}{c}{Est (CI)} & \% & \multicolumn{1}{c}{Est (CI)} & \% & \multicolumn{1}{c}{Est (CI)} & \% & \multicolumn{1}{c}{Est (CI)} & \% & \multicolumn{1}{c}{Est (CI)} & \% \\
    \midrule
       \multirow{2}{*}{Intercept}  & 1 & 2.63 (2.31, 3.40) & -- & 2.47 (2.21, 2.75) & -- & 0.73 (0.42, 1.02) & -- & 0.49 (0.18, 0.59) & -- & 1.36 (1.06, 1.72) & --\\
       & 2 &  2.16 (1.70, 2.49) & -- & 2.93 (2.58, 3.21) & -- & 0.59 (0.32, 0.76) & -- & 0.42 (0.01, 0.76) & -- & 1.51 (1.08, 1.81) & -- \\
       \midrule
       \multirow{2}{*}{Temperature}  & 1 & -0.06 (-0.07, -0.05) & 100 & -0.02 (-0.03, -0.02) & 100 & -0.02 (-0.03, -0.01) & 100 & -0.01 (-0.02, -0.01) & 100 & -0.01 (-0.02, -0.01) & 100 \\
       & 2 &  -0.01 (-0.03, -0.01) & 100 & -0.02 (-0.04, -0.02) & 100 & 0 (0.00, 0.00) & 13 & 0 (-0.01, 0.00) & 49 & -0.02 (-0.03, -0.01) & 100 \\
       \midrule
       \multirow{2}{*}{Weekend} & 1 & -0.83 (-1.03, -0.73) & 100 & -0.28 (-0.39, -0.23) & 100 & 0 (-0.04, 0.00) & 15 & -0.42 (-0.53, -0.35) & 100 & -0.70 (-0.85, -0.63) & 100 \\
       & 2 & -1.05 (-1.22, -0.91) & 100 & -0.66 (-0.77, -0.57) & 100 & 0 (-0.02, 0.03) & 11 & -0.02 (-0.23, 0.00) & 89 & -0.17 (-0.35, -0.08) & 100 \\
       \bottomrule
    \end{tabular}
    }
\end{table}
The temperature effect matches the summer-winter seasonality highlighted above. We can also notice how, regardless of the temperature, weekends exhibit consistently lower average pollution levels -- except for PM$_1$ -- supporting the presence of different traffic patterns between weekdays and weekends that contribute to a reduction in pollution emissions.
The estimates of the auto-regressive coefficients are summarized in Figure \ref{fig:estdvarcoef}, using a similar scheme to that utilized in the simulation study.
We can see how the LASSO penalty shrinks to $0$ most of the auto-regressive coefficients, and it does so consistently across the bootstrap samples. This means that, as expected, the maximum lag $H=7$ was too large and gives rise to very sparse relationships. However, it proves to be very useful as it allows detecting some significant seasonality patterns occurring within or across pollutants at lag $6$ and $7$.

\begin{figure}
    \centering
   \begin{subfigure}[b]{.48\textwidth}
   \centering
   \includegraphics[width=0.99\textwidth]{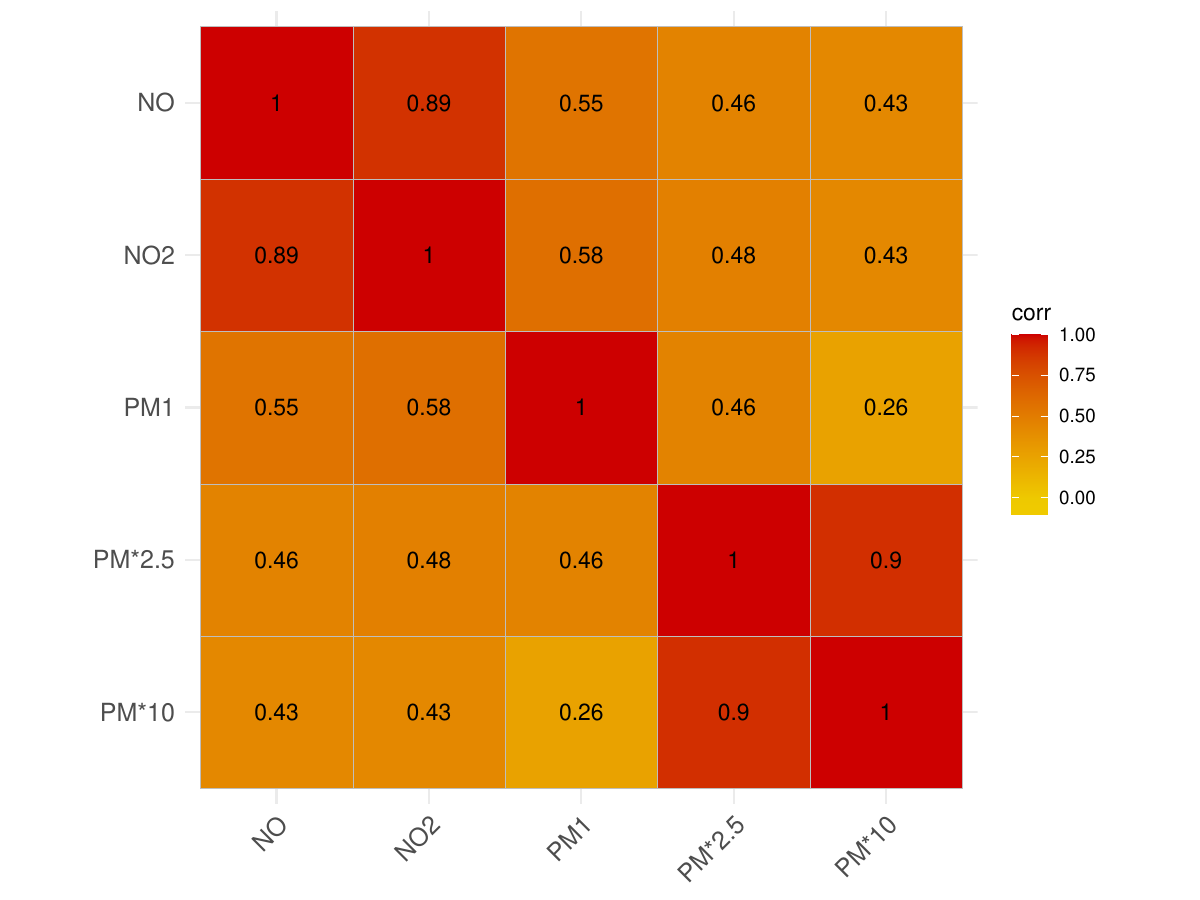}
   \caption{$k = 1$}
   \label{corest1}
   \end{subfigure}
   \begin{subfigure}[b]{.48\textwidth}
   \centering
   \includegraphics[width=0.99\textwidth]{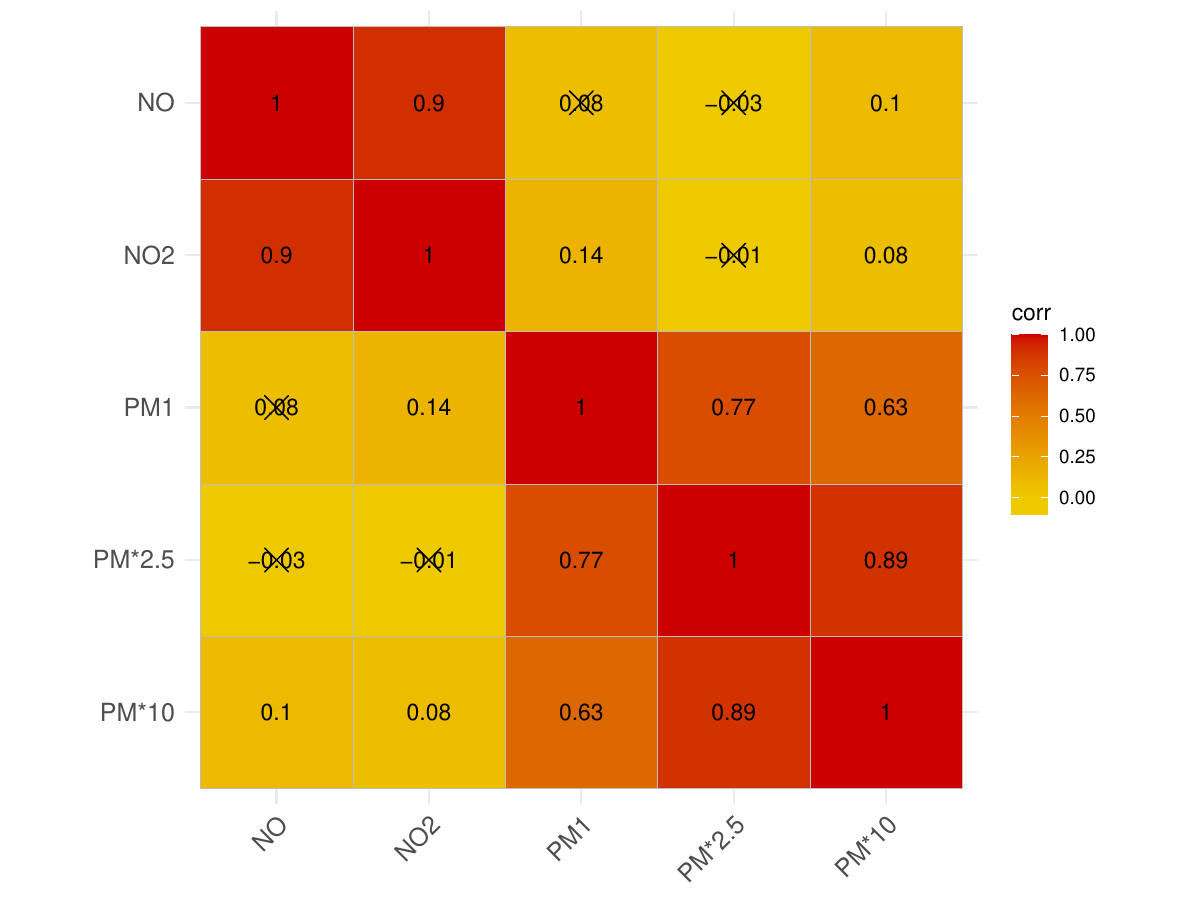}
   \caption{$k = 2$}
   \label{corest2}
   \end{subfigure}
    \caption{Estimated correlation matrix for each state.}
    \label{fig:corest}
\end{figure}

The estimated correlation patterns within each state are reported in Figure \ref{corest1}.
In State 1, pollutants are generally more correlated with each other. More specifically, PM$^*_{2.5}$ and PM$^*_{10}$ show a strong mutual correlation, as do NO and NO$_2$, which are more strongly associated with each other than with other pollutants. PM$_1$, although still correlated, seems to behave as its own cluster. In State 2, corresponding to lower pollution levels, two distinct clusters of highly correlated variables emerge, which are only weakly correlated with each other: nitrogen oxides and particulate matter. This shift reflects changes in dominant emission patterns and atmospheric dynamics. At higher pollution levels, pollutant sources tend to be common and localized. In contrast, at lower pollution levels, differences in source types and atmospheric behavior become more pronounced. Nitrogens primarily originate from combustion processes such as traffic emissions, whereas PM includes both primary particles and secondary aerosols formed from various precursors. Moreover, PM levels are influenced by diffuse and less-regulated sources like residential heating, agriculture, and regional transport, while nitrogens levels are short-lived and more spatially confined. As a result, the correlation between nitrogens and PMs weakens under cleaner atmospheric conditions.

Table \ref{tab:resestdwell} reports the estimated dwell-time regression coefficients and corresponding 95\% bootstrapped confidence intervals. 
\begin{table}
    \centering
     \caption{Estimated parameters of the dwell time: estimates and $95\%$ confidence intervals.}
    \label{tab:resestdwell}
    \begin{tabular}{l|cccc}
    \toprule
    $k$ & $\beta_{0k}$ & $\beta_{1k}$ & Precipitation & Wind speed \\
    \midrule
       1  & -3.35 (-4.41, -2.92) &  -0.12 (-0.18, -0.04) & 6.65 (5.50, 8.64) & 0.49 (0.36, 0.68) \\
       2  & 1.61 (1.00, 2.29) & -0.02 (-0.07, 0.12) & -3.13 (-4.47, -2.14) & -0.46 (-0.71, -0.32) \\
       \bottomrule
    \end{tabular}
\end{table}
State 2 follows a geometric sojourn distribution, as indicated by a non-significant $\beta_{11}$, and is characterized by relatively short durations (at most 5 days). Transitions from this low-pollution state to the high-pollution state become less likely with increasing total precipitation and average wind speed, suggesting that rainy and/or windy conditions prolong cleaner atmospheric episodes. Conversely, in the absence of rain and wind, the probability of switching from high to low pollution decreases over time ($\beta_{12} < 0$), leading to longer persistence in the polluted state—up to 10 days on average (the plot of the estimated distributions is reported in the Supplementary Materials.
This indicates that the polluted state can behave as a \textit{quasi-absorbing} state that is left only when specific meteorological conditions occur. This behavior is consistent with the phenomenon of accumulation, in which stagnant atmospheric conditions inhibit pollutant dispersion, leading to a progressive build-up of emissions from traffic, heating, and industrial activities. Such episodes can persist until wind or precipitation restores vertical and horizontal mixing.

\subsection{Risk Measures with a Focus on Multivariate Risk Assessment}
\label{subsec:risk}
Risk measures have historically been developed in the econometric literature to quantify the systemic risk of financial institutions. In such contexts, risk is typically represented by the realization of particularly low returns. In the context of environmental risk and air pollution, this perspective must be reversed, with risk corresponding to the realization of particularly high concentrations of air pollutants.
Let $\calS=\lcur \text{NO},\dots,\text{PM}^*_{10}\rcur$ the set of $p=5$ pollutants under analysis. For any pollutant $i \in \calS$, we can thereby define the Value-at-Risk $VaR_i(\tau)$ as the minimum log-concentration that we would observe in the $\tau\times 100$ worst occasions and the Expected-Shortfall $ES_i(\tau)$ as the tail-expectation given that the pollutant log-concentration is above the corresponding $VaR_i(\tau)$. Typically, the value of $\tau$ is chosen to be small, for example $\tau = 0.05$. In the multivariate framework considered here, we extend $VaR$ and $ES$ to their multivariate counterparts, $MCoVaR$ and $MCoES$ \citep{tobias2016covar}. The core idea remains unchanged but, instead of examining the marginal distribution of each pollutant $i$, we condition on subsets of the remaining pollutants to be partitioned in two sets $\calH_d\subset\calS\setminus\lcur i\rcur$ (\textit{in-distress} situation) and $\calH_n=\calS\setminus\lcur \calH_d, i\rcur$ (\textit{non-in-distress} situation). More details about the definition of $MCoVaR$ and $MCoES$ are reported in the Supplementary Material. 
Furthermore, our analysis relies on a temporally dynamic model. Consequently, the distribution of pollutants is not time-invariant, necessitating the use of dynamic $MCoVaR$ and $MCoES$ measures at each time point $t$ \citep{bernardi2017multiple}. Specifically, we derive the joint distribution of pollutant log-concentrations at each time $t$ conditional on the past history of the process and marginal with respect to the hidden process. This distribution is given by:
\begin{equation*}
\label{eq:jointposteriortimet}
f\lrnd\by_t\given \by_{(t-H):(t-1)};\, \btheta\rrnd =\sum_{k=1}^K\underbrace{\lrnd\sum_{j=1}^K\sum_{d=1}^m\bar{\alpha}_{t-1}(j, d)\cdot \gamma_{jk}(d)\rrnd}_{\psi_{tk}}\cdot f_{k}\lrnd \by_t \given \by_{(t-H):(t-1)};\, \btheta_k\rrnd.
\end{equation*}
Here, $\bar{\alpha}_{t-1}(j, d)$ denotes the normalized forward probability of being in state $j$ with sojourn $d$ at time $t-1$, and $\gamma_{jk}(d)$ is the transition probability from state $j$ to state $k$. The weights $\psi_{tk}$ are the one-step-ahead filtering probabilities \citep{Cappe_etal2005}, which sum to 1 and act as mixture weights in a $K$-component Gaussian mixture at each time $t$ (Equation \eqref{eq:cond_observation_process}).

The multivariate risk measures for any pollutant $i$ vary depending on the partitioning of the remaining pollutants into $\calH_d \cup \calH_n = \calS \setminus \lcur i \rcur$, leading to a combinatorially large number of possible configurations. 
To obtain a more compact summary and assess the overall contribution of each pollutant to the risk of others, we adopt the \textit{Shapley Value} methodology. This approach allows for the decomposition of total risk among different risk factors by computing their marginal contributions across all possible configurations of distress and non-distress scenarios. In this context, the contribution of pollutant $j$ to the risk of pollutant $i$ is quantified by averaging over all possible combinations of other pollutants being in distress or not, resulting in a Shapley Value $SH_{ti}(j)$ (see the Supplementary Material for details).
Originally developed in game theory to fairly distribute gains (quantified in a common unit of measure) among cooperative players, this approach requires a modification to ensure comparability of $SH_{ti}(\cdot)$ across pollutants. Specifically, variations in the risk measure must be interpreted in relative terms. We therefore propose a standardization of the Shapley Value by the variability of pollutant $i$ at time $t$.
Let $\eta_{ti}(\calH_d)$ denote any multivariate risk measures (e.g. $MCoVaR$ or $MCoES$) for pollutant $i$ at time $t$, given that the subset $\calH_d \subset \calS \setminus \lcur i \rcur$ of institutions is in distress, while $\calS \setminus \calH_d$ is not. Our proposed Shapely Value evaluation is as follows:
\begin{equation*}
    \tilde{Sh}_{ti}(j)=\frac{1}{\sigma_{ti}}\cdot \sum_{\calH\subset\calS\setminus\lcur i, j\rcur}v(\calH)\cdot\lrnd\eta_i\lrnd\calH\cup\lcur j\rcur\rrnd-\eta_i\lrnd\calH\rrnd\rrnd,
\end{equation*}
where $\sigma_{ti}$ is the marginal standard deviation of pollutant $i$ at time $t$.

\begin{figure}
        \centering
            \centering
            \includegraphics[width = .9\textwidth]{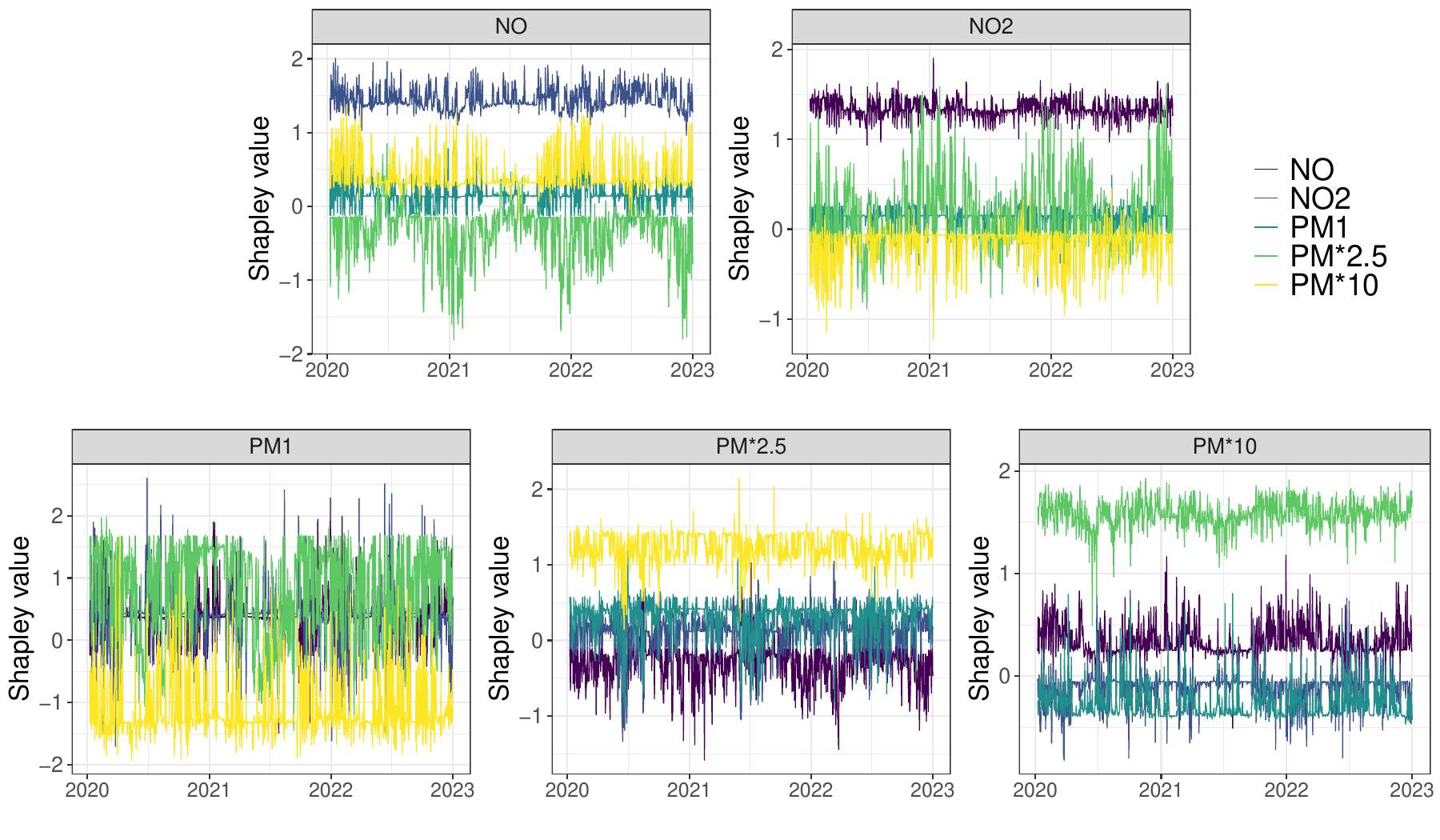}
        \caption{Time series of the Shapley Values of each pollutant on each other for the $MCoVaR$.}
        \label{fig:mcovar}
    \end{figure}

We use this method to evaluate the risk contribution between each pair of pollutant log-concentrations for $\tau=0.05$. The results are reported in Figures \ref{fig:mcovar} for the $MCoVaR$, offering a nuanced view of how each pollutant contributes to the risk of others over time. 
In particular, we observe distinct patterns of interaction and risk attribution. NO and NO$_2$ display consistently high and mutually reinforcing Shapley values. This pattern reflects the strong chemical coupling between these two gases -- NO$_2$ is primarily formed through the oxidation of NO -- and their common emission sources, particularly road traffic. The stable and prominent risk contributions of NO and NO$_2$ indicate that these two pollutants are not only co-occurring but also jointly drive the dynamics of air quality deterioration, particularly in urban and suburban settings. Moreover, the symmetry in their risk profiles highlights the bidirectional nature of their influence.
This mutual influence is especially pronounced during peak pollution episodes, suggesting that risk mitigation strategies targeting one of the two may produce cascading benefits across the system.
In contrast, PM$_1$, PM$^*_{2.5}$, and PM$^*_{10}$ exhibit more heterogeneous patterns. PM$^*_{2.5}$ and PM$^*_{10}$ have high mutual Shapley values, indicating their joint behavior and potential common sources, such as regional particulate matter transport and resuspension. These relationships suggest that the monitoring of one pollutant in each pair may suffice to track the overall risk, reducing the complexity of surveillance systems without substantial loss in informational content.
Conversely, PM$_1$ exhibits a more independent profile, with lower and more variable Shapley values across both risk measures. In some high-pollution periods, PM$_1$ even shows negative contributions to the overall risk of other particulates, suggesting competitive dynamics in particulate composition. Note that a similar interpretation is obtained for the $MCoES$ (see the Supplementary Material for further details). 

\section{Final Discussion}
\label{sec:conc}
We introduced a comprehensive and flexible framework for environmental risk assessment based on a nonhomogeneous hidden semi-Markov model with multivariate, autoregressive, and state-dependent emission structures. The proposed approach represents a novel, data-driven approach to modeling pollution emissions, allowing for the simultaneous estimation of hidden regimes, pollutant dynamics, and risk contributions in a coherent statistical setting.
Simulation studies confirm the reliability of the estimation procedure, demonstrating that the model parameters are accurately recovered under various configurations. Importantly, the computational burden of the estimation process is manageable and scalable. While model complexity increases with the number of hidden states, the length of dwell times, and the dimensionality of the outcome variables, the framework remains tractable and can be implemented efficiently with modern computing resources.
The model’s structure also lends itself to further generalization, incorporating other kinds of penalties to improve regularization or alternative specifications of the hazard function,

Empirical application to real-world air quality data illustrates the interpretability and relevance of the model. 
For instance, the analysis of pollution patterns in Danmarkplass confirms that the air quality is generally clean, though temporary episodes of elevated pollutant concentrations occur during periods of intensified human activity and unfavorable meteorological conditions.
Overall, the proposed model-based approach provides a robust statistical framework for environmental risk assessment, effectively capturing nonstationary, multivariate, and regime-switching behaviors in pollution data. By integrating dynamic risk measures, time-varying dependence structures from a hidden semi-Markov model, and Shapley value decomposition, it also enables interpretable, time-resolved analysis of inter-pollutant risk propagation. This supports more targeted interventions by identifying key risk-driving pollutants and tracking their evolving influence, with broad applicability beyond air quality monitoring.



\section*{Data availability statement}
The air quality data from Danmarksplass, Bergen, is provided by the National Air Quality Reference Laboratory and is publicly available at \url{https://luftkvalitet.nilu.no/en/historical}. The weather observations from station SN50540 Florida, Bergen,  are downloaded from \url{https://seklima.met.no/en} and are managed by The Norwegian Meteorological Institute. 

\section*{Funding}
This work has been supported by MIUR, grant number 2022XRHT8R - The SMILE project: Statistical Modelling and Inference for Living the Environment.
The contribution of Pierfrancesco Alaimo Di Loro to this work has been supported by PON “Ricerca e Innovazione” 2014-2020 (PON R\&I FSE-REACT EU), Azione IV.6 “Contratti di ricerca sutematiche Green,” grant number 60-G-34690-1.

\newpage
\bibliographystyle{abbrvnat} 

\begin{thebibliography}{42}
\providecommand{\natexlab}[1]{#1}
\providecommand{\url}[1]{\texttt{#1}}
\expandafter\ifx\csname urlstyle\endcsname\relax
  \providecommand{\doi}[1]{doi: #1}\else
  \providecommand{\doi}{doi: \begingroup \urlstyle{rm}\Url}\fi

\bibitem[Adrian and Brunnermeier(2016)]{tobias2016covar}
T.~Adrian and M.~K. Brunnermeier.
\newblock Covar.
\newblock \emph{The American Economic Review}, 106\penalty0 (7):\penalty0 1705, 2016.

\bibitem[Andersson(2021)]{andersson2021mechanisms}
A.~Andersson.
\newblock Mechanisms for log normal concentration distributions in the environment.
\newblock \emph{Scientific reports}, 11\penalty0 (1):\penalty0 16418, 2021.

\bibitem[Baraga{\~n}o et~al.(2022)Baraga{\~n}o, Rati{\'e}, Sierra, Chrastn{\`y}, Kom{\'a}rek, and Gallego]{baragano2022multiple}
D.~Baraga{\~n}o, G.~Rati{\'e}, C.~Sierra, V.~Chrastn{\`y}, M.~Kom{\'a}rek, and J.~Gallego.
\newblock Multiple pollution sources unravelled by environmental forensics techniques and multivariate statistics.
\newblock \emph{Journal of hazardous materials}, 424:\penalty0 127413, 2022.

\bibitem[Barbu and Limnios(2009)]{barbu2009semi}
V.~S. Barbu and N.~Limnios.
\newblock \emph{Semi-{M}arkov chains and hidden semi-{M}arkov models toward applications: their use in reliability and DNA analysis}, volume 191.
\newblock Springer Science \& Business Media, 2009.

\bibitem[Basu and Michailidis(2015)]{basu2015highdimensionalvar}
S.~Basu and G.~Michailidis.
\newblock {Regularized estimation in sparse high-dimensional time series models}.
\newblock \emph{The Annals of Statistics}, 43\penalty0 (4):\penalty0 1535 -- 1567, 2015.

\bibitem[{Bergen Kommune}(2023)]{bergen_air_quality_2023}
{Bergen Kommune}.
\newblock {Årsrapport luftkvalitet i Bergen 2023}.
\newblock Technical report, Bergen Kommune, 2023.

\bibitem[Bernardi et~al.(2017)Bernardi, Maruotti, and Petrella]{bernardi2017multiple}
M.~Bernardi, A.~Maruotti, and L.~Petrella.
\newblock Multiple risk measures for multivariate dynamic heavy--tailed models.
\newblock \emph{Journal of Empirical Finance}, 43:\penalty0 1--32, 2017.

\bibitem[Biernacki et~al.(2000)Biernacki, Celeux, and Govaert]{Biernacki2000}
C.~Biernacki, G.~Celeux, and G.~Govaert.
\newblock Assessing a mixture model for clustering with the integrated completed likelihood.
\newblock \emph{IEEE Transactions on Pattern Analysis and Machine Intelligence}, 22\penalty0 (7):\penalty0 719--725, 2000.

\bibitem[Boaz et~al.(2019)Boaz, Lawson, and Pearce]{boaz2019multivariate}
R.~Boaz, A.~Lawson, and J.~Pearce.
\newblock Multivariate air pollution prediction modeling with partial missingness.
\newblock \emph{Environmetrics}, 30\penalty0 (7):\penalty0 e2592, 2019.

\bibitem[Bouveyron et~al.(2022)Bouveyron, Jacques, Schmutz, Simoes, and Bottini]{bouveyron2022co}
C.~Bouveyron, J.~Jacques, A.~Schmutz, F.~Simoes, and S.~Bottini.
\newblock Co-clustering of multivariate functional data for the analysis of air pollution in the south of france.
\newblock \emph{The Annals of Applied Statistics}, 16\penalty0 (3):\penalty0 1400--1422, 2022.

\bibitem[B{\"u}hlmann and Van De~Geer(2011)]{buhlmann2011statistics}
P.~B{\"u}hlmann and S.~Van De~Geer.
\newblock \emph{Statistics for high-dimensional data: methods, theory and applications}.
\newblock Springer Science \& Business Media, 2011.

\bibitem[Cao(2024)]{cao2024integration}
C.~Cao.
\newblock Integration of ten years of daily weather, traffic, and air pollution data from {N}orway’s six largest cities.
\newblock \emph{Scientific Data}, 11\penalty0 (1):\penalty0 744, 2024.

\bibitem[Capp\'{e} et~al.(2005)Capp\'{e}, Moulines, and Ryd\'{e}n]{Cappe_etal2005}
O.~Capp\'{e}, E.~Moulines, and T.~Ryd\'{e}n.
\newblock \emph{Inference in hidden {M}arkov models}.
\newblock Springer, 2005.

\bibitem[Chatterjee and Lahiri(2011)]{chatterjee2011bootstrapping}
A.~Chatterjee and S.~N. Lahiri.
\newblock Bootstrapping lasso estimators.
\newblock \emph{Journal of the American Statistical Association}, 106\penalty0 (494):\penalty0 608--625, 2011.

\bibitem[Efron(2000)]{efron2000bootstrap}
B.~Efron.
\newblock The bootstrap and modern statistics.
\newblock \emph{Journal of the American Statistical Association}, 95\penalty0 (452):\penalty0 1293--1296, 2000.

\bibitem[Finazzi et~al.(2013)Finazzi, Scott, and Fass{\`o}]{finazzi2013model}
F.~Finazzi, E.~M. Scott, and A.~Fass{\`o}.
\newblock A model-based framework for air quality indices and population risk evaluation, with an application to the analysis of scottish air quality data.
\newblock \emph{Journal of the Royal Statistical Society Series C: Applied Statistics}, 62\penalty0 (2):\penalty0 287--308, 2013.

\bibitem[Fong et~al.(2007)Fong, Li, Yau, and Wong]{fong2007mixture}
P.~W. Fong, W.~K. Li, C.~Yau, and C.~S. Wong.
\newblock On a mixture vector autoregressive model.
\newblock \emph{Canadian Journal of Statistics}, 35\penalty0 (1):\penalty0 135--150, 2007.

\bibitem[Greven et~al.(2011)Greven, Dominici, and Zeger]{greven2011approach}
S.~Greven, F.~Dominici, and S.~Zeger.
\newblock An approach to the estimation of chronic air pollution effects using spatio-temporal information.
\newblock \emph{Journal of the American Statistical Association}, 106\penalty0 (494):\penalty0 396--406, 2011.

\bibitem[Hadj-Amar et~al.(2024)Hadj-Amar, Jewson, and Vannucci]{hadj2024bayesian}
B.~Hadj-Amar, J.~Jewson, and M.~Vannucci.
\newblock Bayesian sparse vector autoregressive switching models with application to human gesture phase segmentation.
\newblock \emph{The Annals of Applied Statistics}, 18\penalty0 (3):\penalty0 2511--2531, 2024.

\bibitem[Khalili and Chen(2007)]{khalili2007variable}
A.~Khalili and J.~Chen.
\newblock Variable selection in finite mixture of regression models.
\newblock \emph{Journal of the American Statistical Association}, 102\penalty0 (479):\penalty0 1025--1038, 2007.

\bibitem[Koslik(2025)]{koslik2025hidden}
J.-O. Koslik.
\newblock Hidden semi-{M}arkov models with inhomogeneous state dwell-time distributions.
\newblock \emph{Computational Statistics \& Data Analysis}, 209:\penalty0 108171, 2025.

\bibitem[Lagona and Mingione(2025)]{lagona2025nonhomogeneous}
F.~Lagona and M.~Mingione.
\newblock Nonhomogeneous hidden semi-{M}arkov models for toroidal data.
\newblock \emph{Journal of the Royal Statistical Society Series C: Applied Statistics}, 74\penalty0 (1):\penalty0 142--166, 2025.

\bibitem[Li(2020)]{li2020debiasing}
S.~Li.
\newblock Debiasing the debiased lasso with bootstrap.
\newblock \emph{Electronic Journal of Statistics}, 14:\penalty0 2298–2337, 2020.
\newblock ISSN 935-7524.

\bibitem[Liang et~al.(2021)Liang, Zhang, Chang, and Huang]{liang2021modeling}
D.~Liang, H.~Zhang, X.~Chang, and H.~Huang.
\newblock Modeling and regionalization of {C}hina’s {PM}2.5 using spatial-functional mixture models.
\newblock \emph{Journal of the American Statistical Association}, 116\penalty0 (533):\penalty0 116--132, 2021.

\bibitem[Liao et~al.(2021)Liao, Park, Zhang, Cheng, Ji, Ying, and Yu]{liao2021multiple}
K.~Liao, E.~S. Park, J.~Zhang, L.~Cheng, D.~Ji, Q.~Ying, and J.~Z. Yu.
\newblock A multiple linear regression model with multiplicative log-normal error term for atmospheric concentration data.
\newblock \emph{Science of The Total Environment}, 767:\penalty0 144282, 2021.

\bibitem[Maruotti et~al.(2017)Maruotti, Bulla, Lagona, Picone, and Martella]{maruotti2017dynamic}
A.~Maruotti, J.~Bulla, F.~Lagona, M.~Picone, and F.~Martella.
\newblock Dynamic mixture of factor analyzers to characterize multivariate air pollutant exposures.
\newblock \emph{The Annals of Applied Statistics}, 11\penalty0 (3):\penalty0 1617--1648, 2017.

\bibitem[Mork et~al.(2024)Mork, Kioumourtzoglou, Weisskopf, Coull, and Wilson]{mork2024heterogeneous}
D.~Mork, M.-A. Kioumourtzoglou, M.~Weisskopf, B.~A. Coull, and A.~Wilson.
\newblock Heterogeneous distributed lag models to estimate personalized effects of maternal exposures to air pollution.
\newblock \emph{Journal of the American Statistical Association}, 119\penalty0 (545):\penalty0 14--26, 2024.

\bibitem[Ott(1990)]{ott1990physical}
W.~R. Ott.
\newblock A physical explanation of the lognormality of pollutant concentrations.
\newblock \emph{Journal of the Air \& Waste Management Association}, 40\penalty0 (10):\penalty0 1378--1383, 1990.

\bibitem[Ouyang et~al.(2015)Ouyang, Guo, Cai, Li, Han, Liu, and Liu]{ouyang2015washing}
W.~Ouyang, B.~Guo, G.~Cai, Q.~Li, S.~Han, B.~Liu, and X.~Liu.
\newblock The washing effect of precipitation on particulate matter and the pollution dynamics of rainwater in downtown beijing.
\newblock \emph{Science of the Total Environment}, 505:\penalty0 306--314, 2015.

\bibitem[O’Connell and Højsgaard(2011)]{mhsmm_package}
J.~O’Connell and S.~Højsgaard.
\newblock Hidden semi {M}arkov models for multiple observation sequences: The mhsmm package for r.
\newblock \emph{Journal of Statistical Software}, 39:\penalty0 1--22, 2011.

\bibitem[Raymaekers and Rousseeuw(2024)]{raymaekers2024cellwise}
J.~Raymaekers and P.~J. Rousseeuw.
\newblock The cellwise minimum covariance determinant estimator.
\newblock \emph{Journal of the American Statistical Association}, 119\penalty0 (548):\penalty0 2610--2621, 2024.

\bibitem[Ricciotti et~al.(2025)Ricciotti, Picone, Pollice, and Maruotti]{ricciotti2025zero}
L.~Ricciotti, M.~Picone, A.~Pollice, and A.~Maruotti.
\newblock A zero-inflated hidden semi-{M}arkov model with covariate-dependent sojourn parameters for analysing marine data in the venice lagoon.
\newblock \emph{Journal of the Royal Statistical Society Series C: Applied Statistics}, 74\penalty0 (2):\penalty0 506--529, 2025.

\bibitem[Ruiz-Suarez et~al.(2022)Ruiz-Suarez, Leos-Barajas, and Morales]{ruiz2022hidden}
S.~Ruiz-Suarez, V.~Leos-Barajas, and J.~M. Morales.
\newblock Hidden {M}arkov and semi-{M}arkov models when and why are these models useful for classifying states in time series data?
\newblock \emph{Journal of Agricultural, Biological and Environmental Statistics}, pages 1--25, 2022.

\bibitem[Shan et~al.(2024)Shan, Casey, Shearston, and Henneman]{shan2024methods}
X.~Shan, J.~A. Casey, J.~A. Shearston, and L.~R. Henneman.
\newblock Methods for quantifying source-specific air pollution exposure to serve epidemiology, risk assessment, and environmental justice.
\newblock \emph{GeoHealth}, 8\penalty0 (11):\penalty0 e2024GH001188, 2024.

\bibitem[St{\"a}dler and Mukherjee(2013)]{stadler2013penalized}
N.~St{\"a}dler and S.~Mukherjee.
\newblock {Penalized estimation in high-dimensional hidden {M}arkov models with state-specific graphical models}.
\newblock \emph{The Annals of Applied Statistics}, pages 2157--2179, 2013.

\bibitem[St{\"a}dler et~al.(2010)St{\"a}dler, B{\"u}hlmann, and Van De~Geer]{stadler2010}
N.~St{\"a}dler, P.~B{\"u}hlmann, and S.~Van De~Geer.
\newblock $\ell$ 1-penalization for mixture regression models.
\newblock \emph{Test}, 19:\penalty0 209--256, 2010.

\bibitem[Tan et~al.(2021)Tan, Chiong, and and]{Tan21102021}
L.~Tan, K.~X. Chiong, and H.~R.~M. and.
\newblock Estimation of high-dimensional seemingly unrelated regression models.
\newblock \emph{Econometric Reviews}, 40\penalty0 (9):\penalty0 830--851, 2021.

\bibitem[Tavella et~al.(2023)Tavella, {Galeao da Rosa Moraes}, {Maciel Aick}, Ramires, Pereira, Soares, and {da Silva Júnior}]{TAVELLA2023101662}
R.~A. Tavella, N.~{Galeao da Rosa Moraes}, C.~D. {Maciel Aick}, P.~F. Ramires, N.~Pereira, A.~G. Soares, and F.~M.~R. {da Silva Júnior}.
\newblock Weekend effect of air pollutants in small and medium-sized cities: The role of policies stringency to covid-19 containment.
\newblock \emph{Atmospheric Pollution Research}, 14\penalty0 (2):\penalty0 101662, 2023.
\newblock ISSN 1309-1042.

\bibitem[Yu(2015)]{yu2015hidden}
S.-Z. Yu.
\newblock \emph{Hidden Semi-{M}arkov models: theory, algorithms and applications}.
\newblock Morgan Kaufmann, 2015.

\bibitem[Zhang et~al.(2018)Zhang, Jiao, Xu, Zhao, Tang, Zhou, and Gong]{zhang2018influences}
B.~Zhang, L.~Jiao, G.~Xu, S.~Zhao, X.~Tang, Y.~Zhou, and C.~Gong.
\newblock Influences of wind and precipitation on different-sized particulate matter concentrations (pm 2.5, pm 10, pm 2.5--10).
\newblock \emph{Meteorology and Atmospheric Physics}, 130:\penalty0 383--392, 2018.

\bibitem[Zhu et~al.(2024)Zhu, Wen, Cao, He, and Wang]{zhu2024review}
G.~Zhu, Y.~Wen, K.~Cao, S.~He, and T.~Wang.
\newblock A review of common statistical methods for dealing with multiple pollutant mixtures and multiple exposures.
\newblock \emph{Frontiers in Public Health}, 12:\penalty0 1377685, 2024.

\bibitem[Zucchini et~al.(2016)Zucchini, MacDonald, and Langrock]{zucchini2009hidden}
W.~Zucchini, I.~L. MacDonald, and R.~Langrock.
\newblock \emph{Hidden {M}arkov models for time series: an introduction using R}.
\newblock Chapman and Hall/CRC, 2016.

\end{thebibliography}

\newpage
\section*{Supplementary information}
\section*{Data description: additional insights}

\begin{figure}
        \centering
        \begin{subfigure}[b]{.3\textwidth}
            \centering
            \includegraphics[width = .95\textwidth]{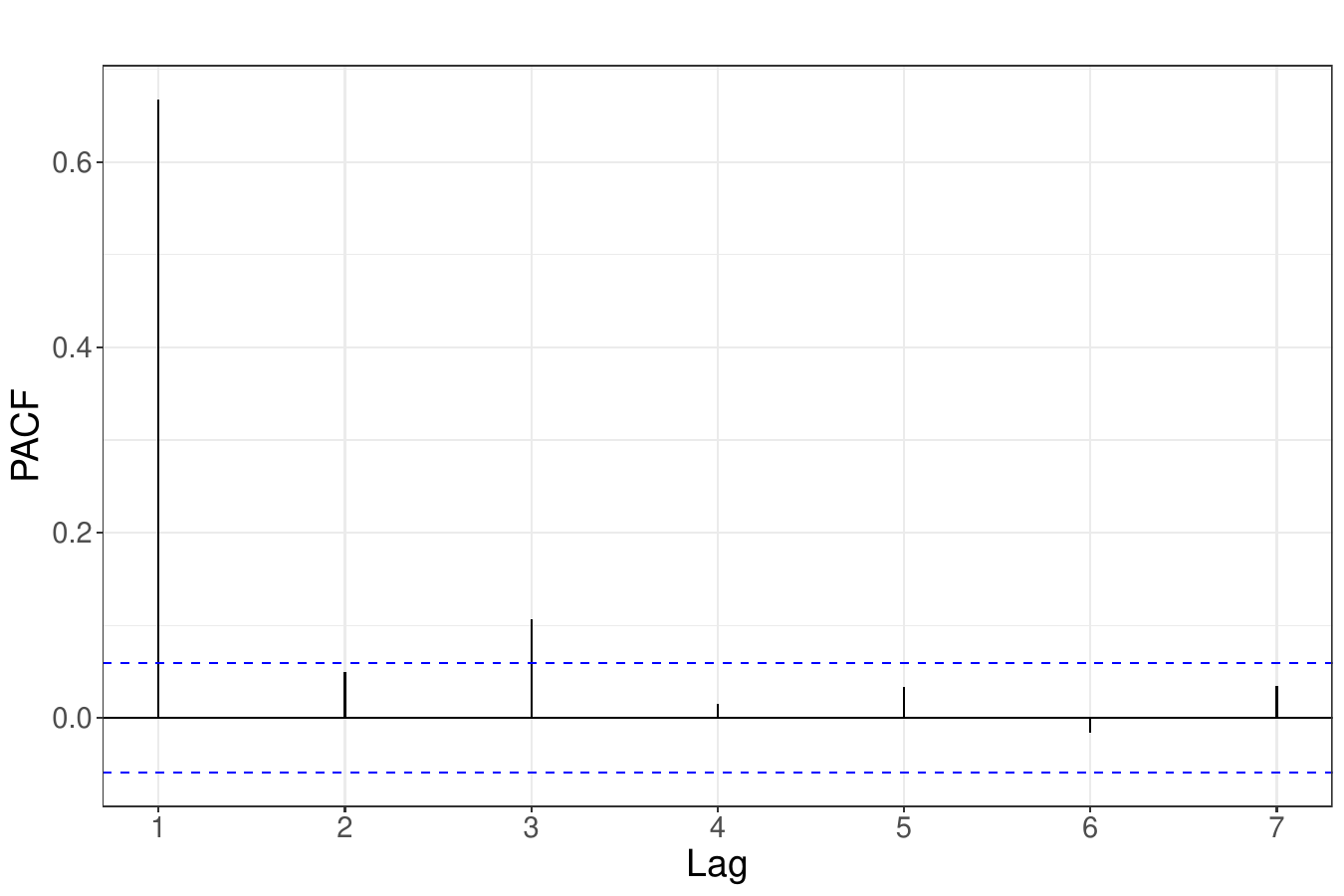}
            \caption*{\scriptsize PM$_1$}
        \end{subfigure}
        \begin{subfigure}[b]{.3\textwidth}
            \centering
            \includegraphics[width = .95\textwidth]{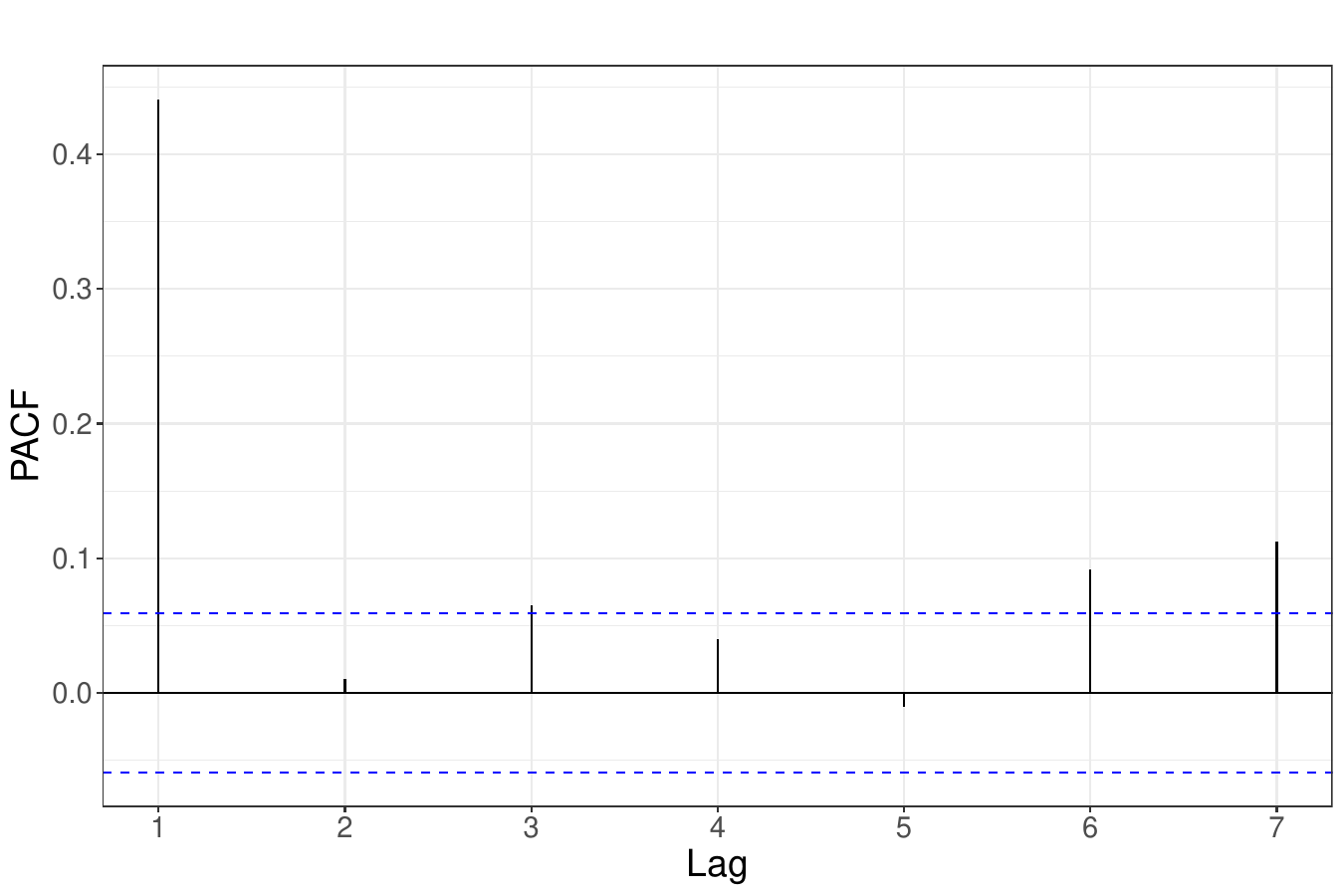}
            \caption*{\scriptsize PM$^*_{2.5}$}
        \end{subfigure}
        \begin{subfigure}[b]{.3\textwidth}
            \centering
            \includegraphics[width = .95\textwidth]{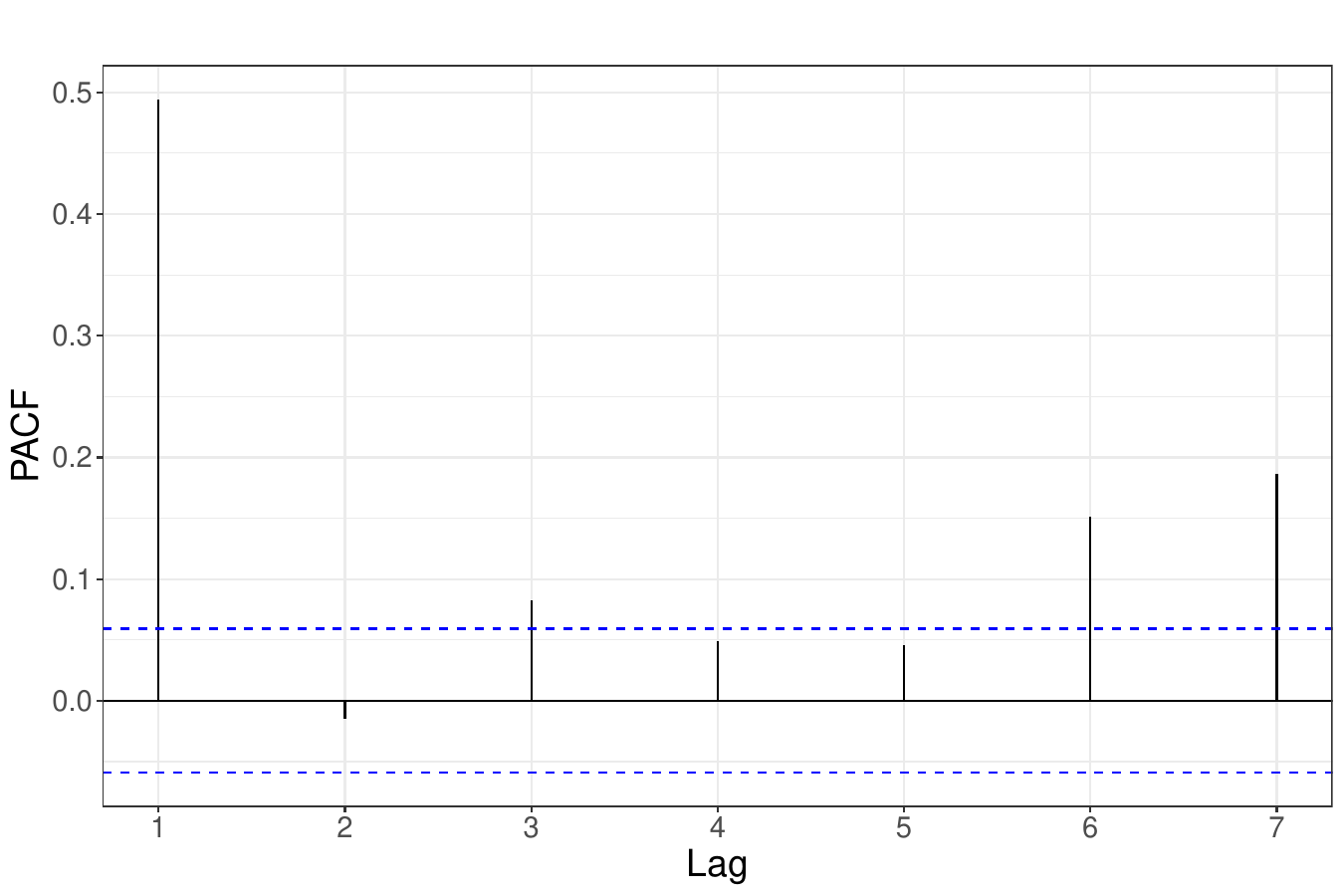}
            \caption*{\scriptsize PM$^*_{10}$}
        \end{subfigure}
        \begin{subfigure}[b]{.3\textwidth}
            \centering
            \includegraphics[width = .95\textwidth]{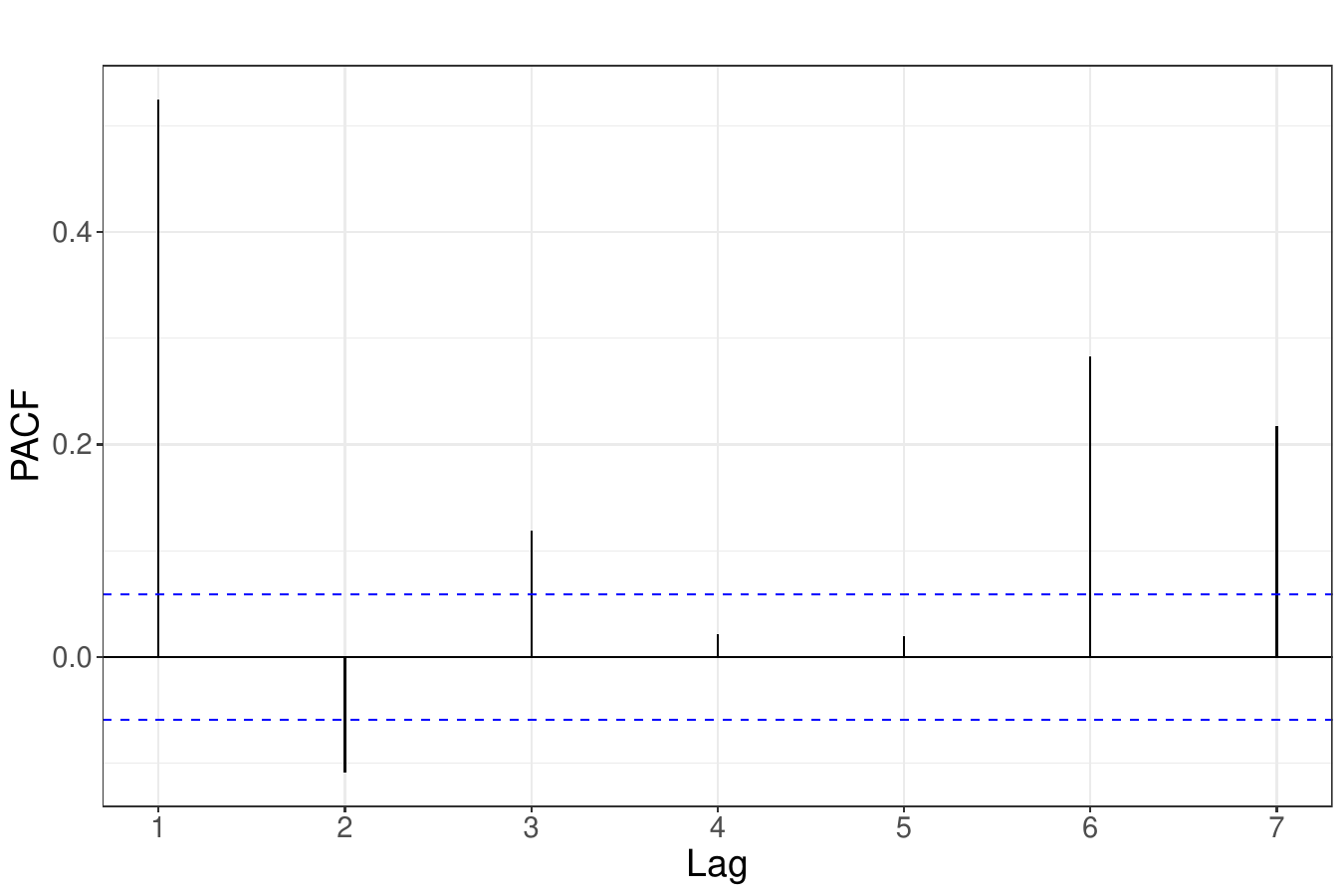}
            \caption*{\scriptsize NO}
        \end{subfigure}
        \begin{subfigure}[b]{.3\textwidth}
            \centering
            \includegraphics[width = .95\textwidth]{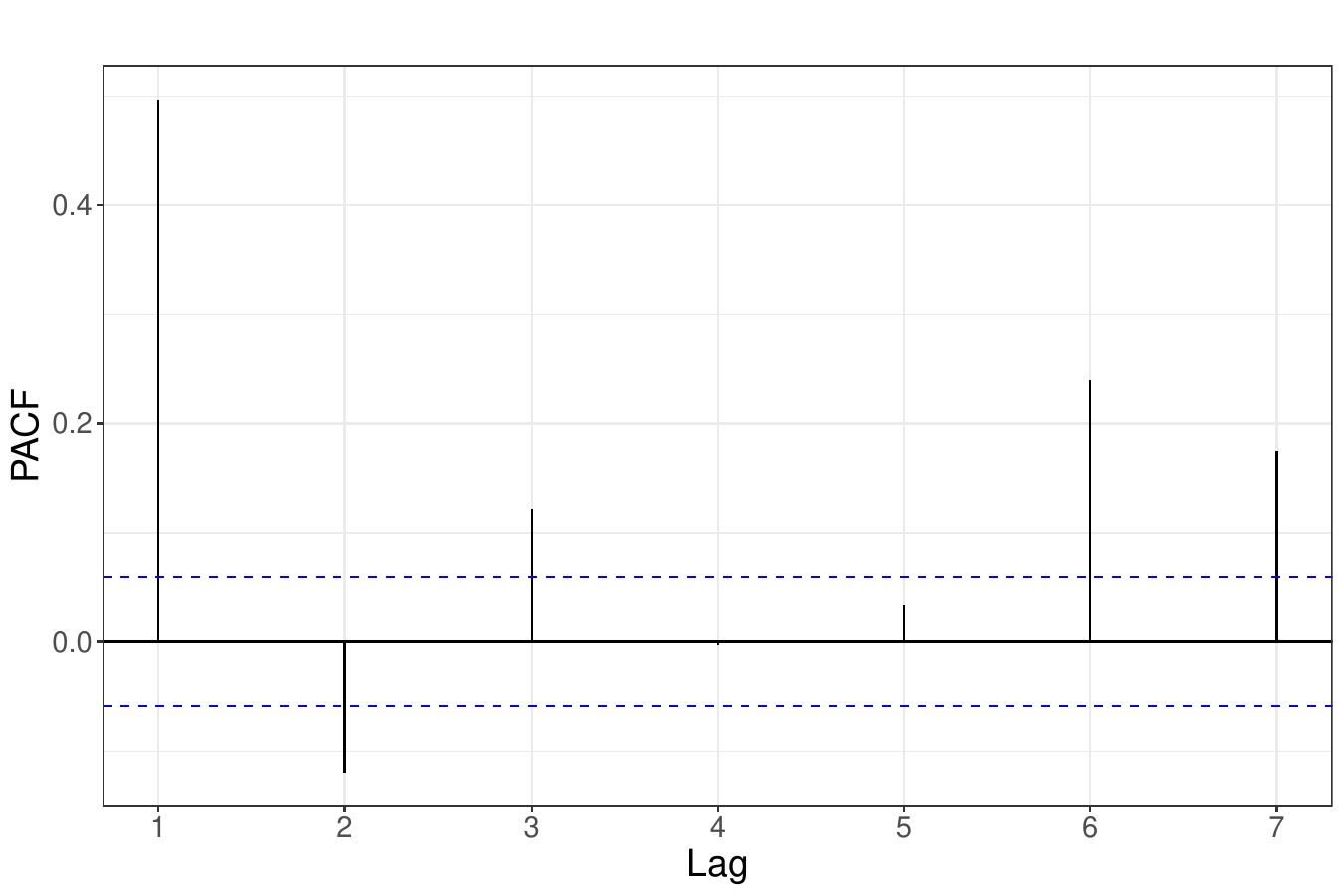}
            \caption*{\scriptsize NO$_2$}
        \end{subfigure}
        \caption{Observed partial autocorrelation functions of the five considered pollutants (on the log-scale): (a) PM$_1$, (b) PM$^*_{2.5}$,(c) PM$^*_{10}$, (d) NO, (e) NO$_2$.}
        \label{fig:obspacf}
    \end{figure}

\begin{figure}[t]
    \centering
    \includegraphics[width=0.9\linewidth]{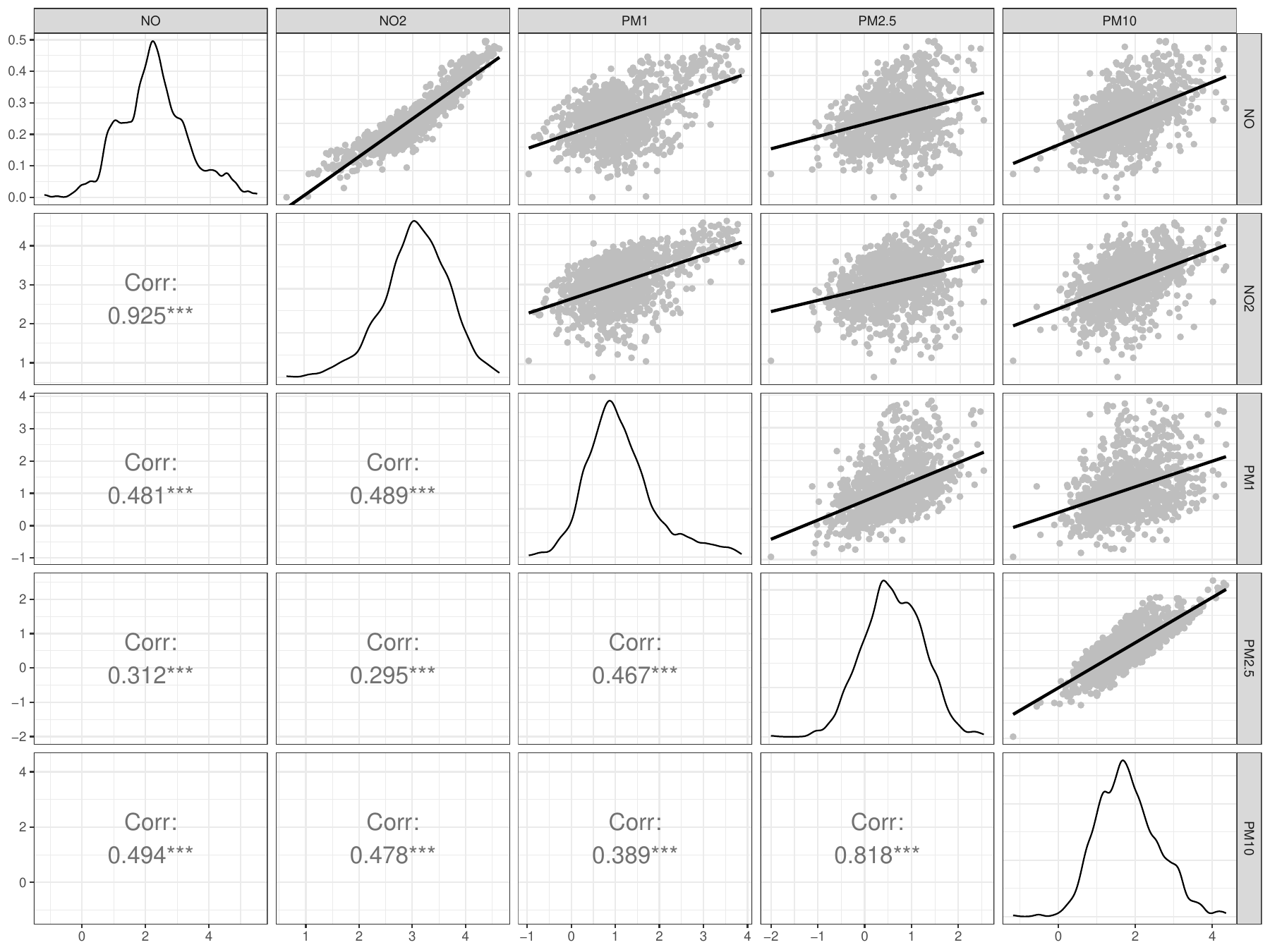}
    \caption{Marginal distribution of each pollutant (diagonal) and pairwise scatterplots with related correlations (*** for statistical significance at $\alpha = 0.01)$.}
    \label{fig:obsdata}
\end{figure}

In this Section, we report some additional descriptive insights on the available data, both in terms of the outcomes and the covariates.
Figure \ref{fig:obspacf} reports the partial auto-correlations functions up to lag $7$ for all log-concentrations of pollutants.
Most lags exhibit non significant partial correlations, but others result to be significant. This happens even for unexpected lags, such as $3$, $6$, and $7$ on the NO2, exhacerbating the need for some automatic lag selection procedure that cannot be driven by prior knowledge or theoretical results about the way pollutants interact through time in an open-environment.
Figure \ref{fig:obsdata} reports all marginal distributions, scatterplos and correlations across the log-concentrations of pollutants.
The marginal distributions of the pollutants are approximately bell-shaped with rapidly decaying tails. However, deviations such as sharp peaks in the NO and $\mathrm{PM}_{2.5}$ distributions indicate possible heterogeneity. 
All pairwise scatterplots display positive correlations, ranging from moderate ($\approx 0.295$) to strong ($\approx 0.925$).

\section*{Simulation study: additional results}

\begin{figure}
    \centering
    \includegraphics[width=0.75\textwidth]{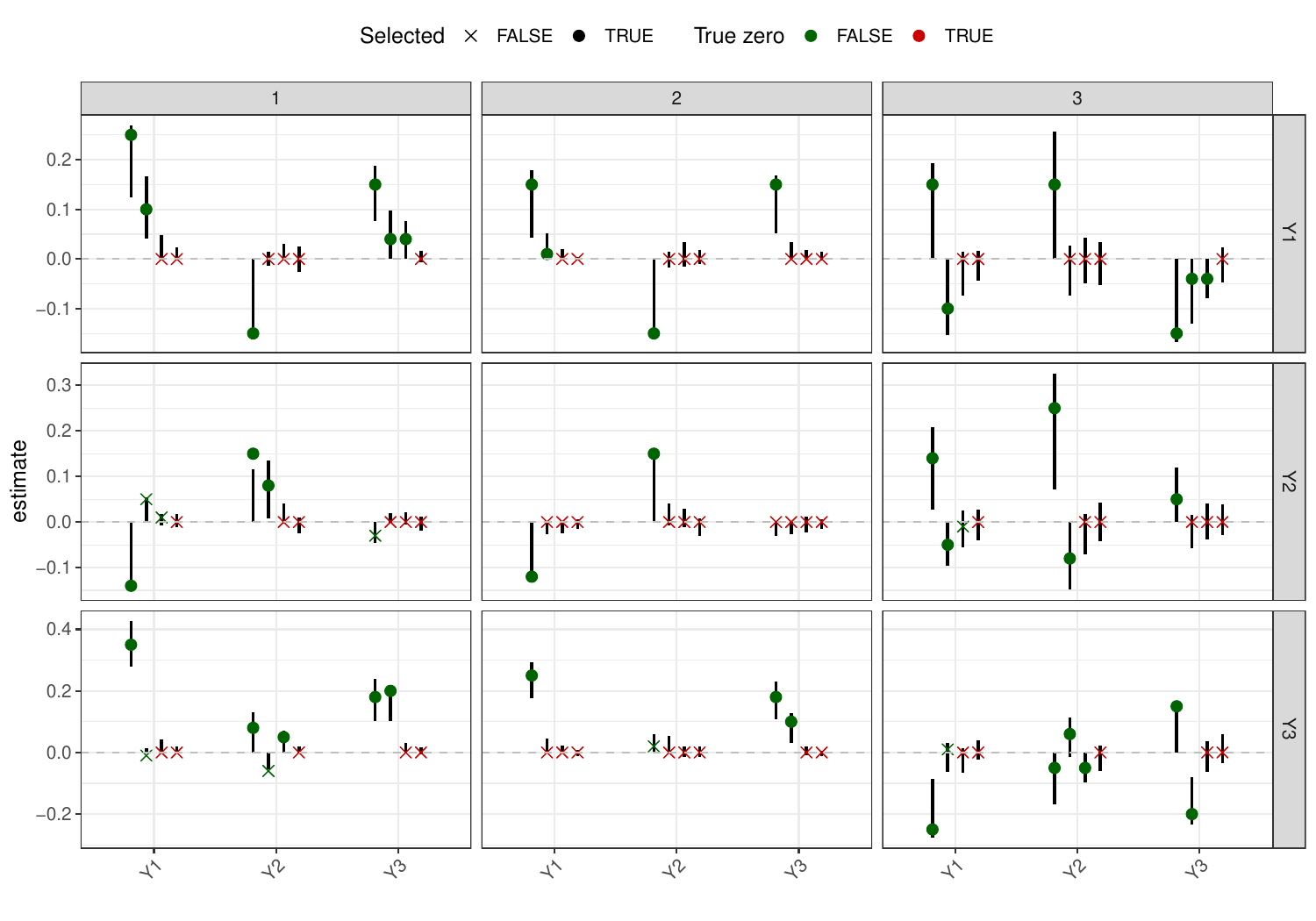}
    \includegraphics[width=0.75\textwidth]{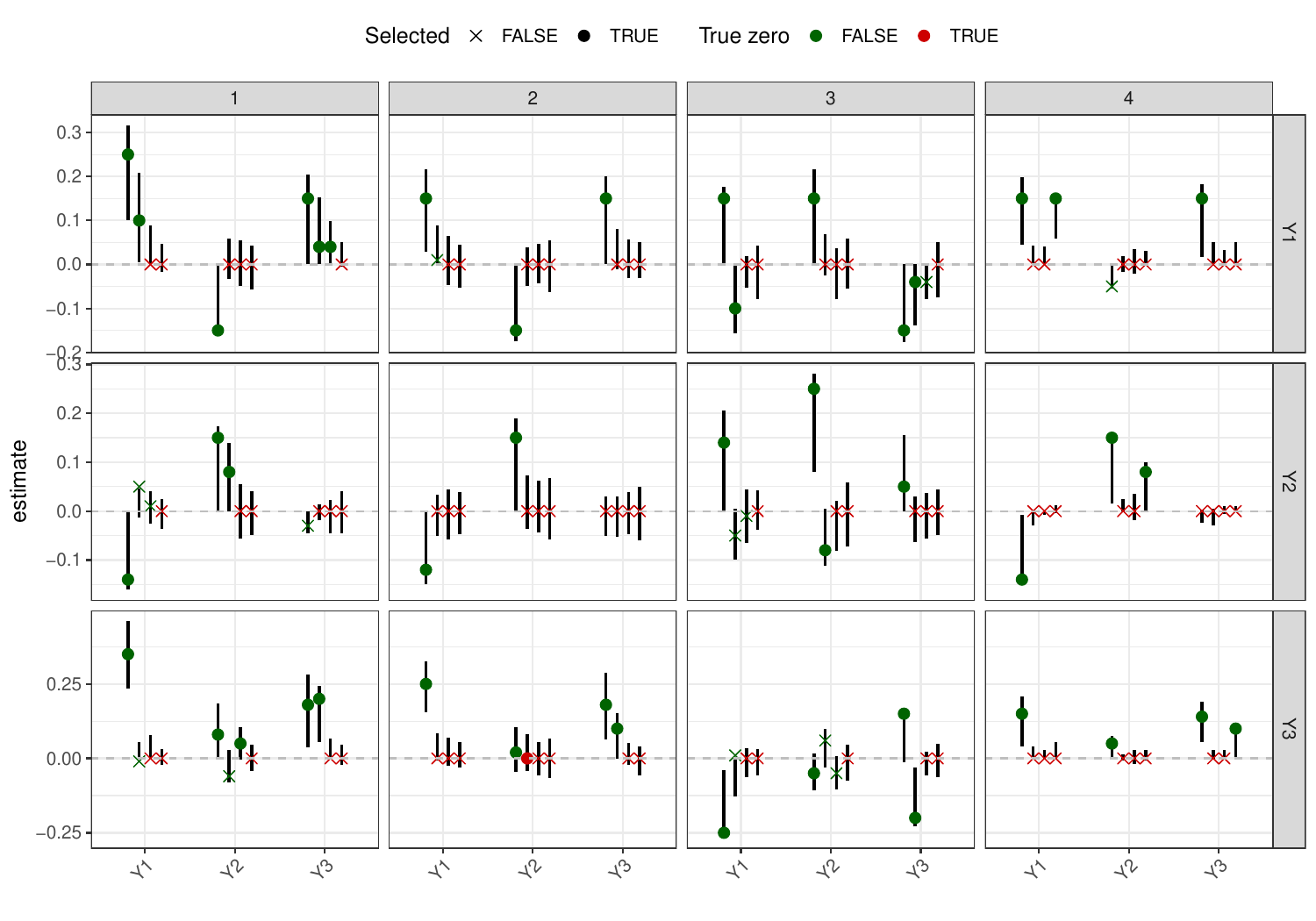}
    \caption{Recovery of the autoregressive coefficients for $K = 3$ (top) and $K=4$ (bottom). True values are: green if different from $0$, red if not; marked as a cross if shrunk to $0$ more than $50\%$ of the times and as a dot otherwise. The black line represents the central $95\%$ of the estimates distribution.}
    \label{fig:sim_VAR34}
\end{figure}

We here report some additional results of the simulation study that are referred in the main text.
Table \ref{tab:simresbetaVAR34} shows the performances of the model in recovering the \textit{true} values of the exogenous covariate impacting on each outcome. 
\begin{table}
    \centering
        \caption{Recovery of the regression coefficients for $K=\tilde{K}^*=3, 4$: True value, Mean of the estimates, $95\%$ central interval, Root Mean Squared Error (RMSE).}
    \adjustbox{max width = .95\textwidth}{%
    \begin{tabular}{lc|ccccc|ccccc|ccccc|ccccc}
    \toprule
      & $k$  & \multicolumn{5}{c}{$1$} & \multicolumn{5}{c}{$2$} & \multicolumn{5}{c}{$3$} & \multicolumn{5}{c}{$4$}\\
      & & True & Mean & CI$_{0.95}$ & RMSE &  \% & True & Mean & CI$_{0.95}$ & RMSE &  \% & True & Mean & CI$_{0.95}$ & RMSE &  \% & True &  Mean & CI$_{0.95}$ &RMSE &  \% \\
       \midrule
       \multirow{6}{*}{$K=3$} & $b_{01}$ & 3 & 3.11 & (2.95, 3.29) & 0.137 & -- & -1 & -0.914 & (-1.06, -0.78) & 0.112 & -- & 0.75 & 0.820 & (0.59, 1.06) & 0.134 & -- & -- & -- & -- & -- & -- \\
       & $b_{02}$ & 1.5 & 1.38 & (1.23, 1.54) & 0.149 & -- & -2 & -2.09 & (-2.26, -1.96) & 0.122 & -- & 0.25 & 0.28 & (0.06, 0.55) & 0.125 & -- & -- & -- & -- & --& --\\
       & $b_{03}$ & 2 & 2.02 & (1.88, 2.16) & 0.077 & -- & -1.5 & -1.50 & (-1.62, -1.40) & 0.058 & -- & -0.25 & -0.31 & (-0.48, -0.09) & 0.122 & -- & -- & -- & -- & -- & --\\
       \cline{2-17}
        & $b_{11}$ & 0.5 & 0.439 & (0.346, 0.522) & 0.08 & 1 & -0.2 & 0.142 & (-0.248, -0.053) & 0.08 & 1 & 0.6 & 0.561 & (0.407, 0.718) & 0.08 & 1 & -- & -- & -- & -- & -- \\
       & $b_{21}$ & 0 & 0.001 & (-0.027, 0.047) & 0.01 & 0.25 & 0.4 & 0.337 & (0.231, 0.426) & 0.08 & 1 & 0 & 0.003 & (-0.096, 0.117) & 0.05 & 0.47 & -- & -- & -- & --& --\\
       & $b_{31}$ & 0 & -0.002 & (-0.050, 0.025) & 0.02 & 0.20 & -0.1 & -0.044 & (-0.129, 0.000) & 0.07 & 0.79 & 0.1 & 0.062 & (-0.016, 0.228) & 0.07 & 0.73 & -- & -- & -- & -- & --\\
       \midrule
       \multirow{6}{*}{$K=4$} & $b_{01}$ & 3 & 3.14 & (2.88, 3.39) & 0.193 & -- & -1 & -0.95 & (-1.19, -0.70) & 0.127 & -- & 0.75 & 0.82 & (0.53, 1.10) & 0.167 & -- & 1.25 & 1.45 & (1.24, 1.66) & 0.231 & -- \\
       & $b_{02}$ & 1.5 & 1.35 & (1.09, 1.63) & 0.208 & -- & -2 & -2.06 & (-2.28, -1.84) & 0.133 & -- & 0.25 & 0.28 & (-0.01, 0.54) & 0.137 & -- & -1 & -1.20 & (-1.44, -0.97) & 0.239 & -- \\
       & $b_{03}$ & 2 & 2.03 & (1.78, 2.29) & 0.143 & -- & -1.5 & -1.51 & (-1.73, -1.29) & 0.118 & -- & -0.25 & -0.37 & (-0.63, -0.16) & 0.180 & -- & 0.5 & 0.57 & (0.41, 0.75) & 0.116 & --\\
       \cline{2-22}
        & $b_{11}$ & 0.5 & 0.441 & (0.294, 0.567) & 0.09 & 1 & -0.2 & -0.148 & (-0.306, -0.002) & 0.09 & 0.98 & 0.6 & 0.521 & (0.371, 0.685) & 0.11 & 1 & 0 & 0.000 & (-0.029, 0.028) & 0.01 & 0.13 \\
       & $b_{21}$ & 0 & -0.006 & (-0.099, 0.056) & 0.04 & 0.38 & 0.4 & 0.350 & (0.216, 0.533) & 0.09 & 1 & 0 & -0.001 & (-0.147, 0.111) & 0.05 & 0.45 & -0.5 & -0.423 & (-0.524, -0.311) & 0.09 & 1 \\
       & $b_{31}$ & 0 & -0.006 & (-0.117, 0.054) & 0.04 & 0.34 & -0.1 & -0.052 & (-0.182, 0.000) & 0.07 & 0.70 & 0.1 & 0.042 & (-0.019, 0.184) & 0.08 & 0.64 & 0 & -0.002 & (-0.050, 0.034) & 0.02 & 0.20\\
         \bottomrule         
    \end{tabular}
    }
    \label{tab:simresbetaVAR34}
\end{table}
We notice how the RMSEs are generally very low and the non-zero and zero coefficients are correctly selected across all scenarios. 
The estimation and selection of the auto-regressive coefficients for $K=3, 4$ are reported in Figure \ref{fig:sim_VAR34}.
Considering the results presented in the main text as well, we observe that performance remains quite consistent across the three scenarios examined. There is some indication of slight over-shrinking in certain states (specifically, states 1 and 3) under the $K = 4$ setting, where a few coefficients are selected in fewer than 50\% of the simulations, despite being truly non-zero. However, their inclusion probabilities are only slightly below the threshold, which is not a major concern. In addition, the assumption that a single baseline $\lambda_0$ is adequate for all states becomes increasingly tenuous as the number of states increases. Each state exhibits unique characteristics and may require individualized adjustments beyond effective sample size to achieve optimality. This issue will be further explored and addressed in future work.
The discrepancy between the true and estimated variance-covariance matrices are evaluated in terms of the KL-discrepancy \citep{raymaekers2024cellwise}. The average values across the $B=300$ simulations are reported in Table \ref{tab:simest_sigma}.
\begin{table}
    \centering
    \caption{Recovery of the variance-covariance matrices: average KL discrepancy.}
    \label{tab:simest_sigma}
    \begin{tabular}{ll|cccc}
    \toprule
        &    & \multicolumn{4}{c}{$KL\lrnd\Sigma_k, \hat{\Sigma}_k\rrnd$}\\
        & k & 1 & 2 & 3 & 4\\
        \midrule
         \multirow{3}{*}{$K$}& 2 & 0.013 & 0.010 & -- & -- \\
         & 3 & 0.020 & 0.016 & 0.036 & -- \\
         & 4 & 0.079 & 0.033 & 0.058 & 0.015 \\
         \bottomrule
    \end{tabular}
\end{table}
We notice how the KL discrepancy are all low and seem to increase as $K$ increases. Considering that the number of states is increasing with fixed number of observations $T$, this might be explained by the fact that each estimate is obtained with a smaller number of observations.
Table \ref{tab:simresbeta34} reports the estimation performances of the conditional dwell-time regression coefficients for $K=3$ and $K=4$.
\begin{table}[t]
    \caption{Recovery of the dwell-time regression coefficients $K=\tilde{K}^*=2$: True value, Mean of the estimates, $95\%$ central interval, Root Mean Squared Error (RMSE).}
    \centering
    \adjustbox{max width=0.95\textwidth}{%
    \begin{tabular}{lc|cccc|cccc|cccc|cccc}
    \toprule
       & $k$ & \multicolumn{4}{c}{1} & \multicolumn{4}{c}{2} & \multicolumn{4}{c}{3} & \multicolumn{4}{c}{4} \\
       & & True & Mean & CI$_{0.95}$ & RMSE & True & Mean & CI$_{0.95}$ & RMSE & True & Mean & CI$_{0.95}$ & RMSE & True & Mean & CI$_{0.95}$ & RMSE \\
       \midrule
       \multirow{3}{*}{$K=3$} & $\beta_0$ & -1 & -1.02 & (-1.32, -0.71) & 0.15 & -2 & -2.05 & (-2.42, -1.66) & 0.22 & -0.50 & -0.56 & (-1.04, -0.17) & 0.23 & -- & -- & -- & --\\
       & $\beta_1$ & 0.15 & 0.16 & (0.07, 0.26) & 0.05  & 0.35 & 0.36 & (0.25, 0.48) & 0.06 & 0 & 0.02 & (-0.09, 0.21) & 0.08 & -- & -- & -- & --\\
       & $\beta_2$ & -0.50 & -0.51 & (-0.69, -0.34) & 0.09 & 0.50 & 0.51 & (0.35, 0.70) & 0.09 & 0.10 & 0.10 & (-0.12, 0.35) & 0.13 & -- & -- & -- & --\\
        \midrule
       \multirow{3}{*}{$K=4$} & $\beta_0$ & -1 & -1.08 & (-1.62, -0.60) & 0.26 & -2 & -2.09 & (-2.76, -1.40) & 0.38 & -0.50 & -0.62 & (-1.20, -0.13) & 0.31 & -3 & -3.07 & (-3.92,  -2.40) & 0.41\\
       & $\beta_1$ & 0.15 & 0.17 & (0.04, 0.32) & 0.08 & 0.35 & 0.38 & (0.19, 0.59) & 0.10 & 0 & 0.03 & (-0.10, 0.23) & 0.10 & 0.50 & 0.51 & (0.36, 0.71) & 0.09\\
       & $\beta_2$ & -0.50 & -0.51 & (-0.73, -0.31) & 0.12 & 0.50 & 0.52 & (0.26, 0.79) & 0.13 & 0.10 & 0.10 & (-0.16, 0.34) & 0.13 & -0.20 & -0.20 & (-0.49, 0.01) & 0.13\\
         \bottomrule         
    \end{tabular}
    }
    \label{tab:simresbeta34}
\end{table}
Table \ref{tab:simest_omega} reports the estimation performances of the conditional transition probability matrix elements for $K=3$ and $K=4$.
\begin{table}
\centering
\caption{Recovery of the conditional transition probabilities for $K=\tilde{K}^*=3, 4$: True value, Mean of the estimates, $95\%$ central interval, Root Mean Squared Error (RMSE).}
\label{tab:simest_omega}
\begin{tabular}{l|ccc|ccc}
\toprule%
\multirow{2}{*}{$\bOm$}&  \multicolumn{3}{c}{$K=3$} & 
\multicolumn{3}{c}{$K=4$} \\
\cline{2-4}\cline{5-7}%
&  True & Mean & CI & True & Mean & CI  \\
\midrule
$\omega_{12}$ &  0.50 & 0.505 & (0.420, 0.586) & 0.25 & 0.246 & (0.153, 0.329) \\
$\omega_{13}$ & 0.50 & 0.495 & (0.414, 0.580) & 0.25 & 0.252 & (0.159, 0.346) \\
 $\omega_{14}$ & -- & -- & -- & 0.50 & 0.502 & (0.404, 0.607) \\
 $\omega_{21}$ & 0.90 & 0.897 & (0.802, 0.962) &  0.70 & 0.699 & (0.558, 0.822) \\
$\omega_{23}$ & 0.10 & 0.103 & (0.038, 0.198) & 0.20 & 0.198 & (0.068, 0.356) \\
$\omega_{24}$ & -- & -- & -- &  0.10 & 0.103 & (0.016, 0.208) \\
 $\omega_{31}$ & 0.45 & 0.449 & (0.357, 0.551) & 0.15 & 0.146 & (0.060, 0.241) \\
 $\omega_{32}$ &  0.55 & 0.551 & (0.449, 0.643) & 0.25 & 0.254 & (0.173, 0.365) \\
 $\omega_{34}$ & -- & -- & -- & 0.60 & 0.600 & (0.495, 0.693) \\
$\omega_{41}$ & -- & -- & -- & 0.30 & 0.306 & (0.205, 0.393) \\
 $\omega_{42}$ & -- & -- & -- & 0.20 & 0.205 & (0.105, 0.317) \\
 $\omega_{43}$ & -- & -- & -- & 0.50 & 0.489 & (0.376, 0.592) \\
\bottomrule
\end{tabular}
\end{table}
The estimates are very close to the true values in all settings, with errors at the third decimal digit. In all cases, the true value is included in the $95\%$ central interval of the simulated sets.
Finally, Table \ref{tab:simaccuracy} reports a detail of the performances in the segmentation of the time-series into the latent states. 
\begin{table}
    \centering
    \caption{Average ARI and accuracy of the MAP estimates of the latent states}
    \label{tab:simaccuracy}
    \begin{tabular}{l|ccc}
    \toprule
    $K$ & 2 & 3 & 4 \\
\midrule
       ARI  &  0.999 & 0.939 & 0.867 \\
       Accuracy & 0.999 & 0.976 & 0.949 \\
       \bottomrule
    \end{tabular}
\end{table}
ARI and Accuracy are both very high, with the former ranging from $86.7\%$ in the $K=4$ case to $99.9\%$ in the $K=2$ one and the latter ranging from $94.9\%$ in the $K=4$ case to $100\%$ in the $K=2$ one. The degrading performances for increasing $K$ are not related to its size per-sè, but to the fact that the first two components are quite well-separated while the third and the fourth are overlapping with the original ones.

\section*{Real Data application: additional results}

\begin{figure}
    \centering
    \includegraphics[width=0.7\linewidth]{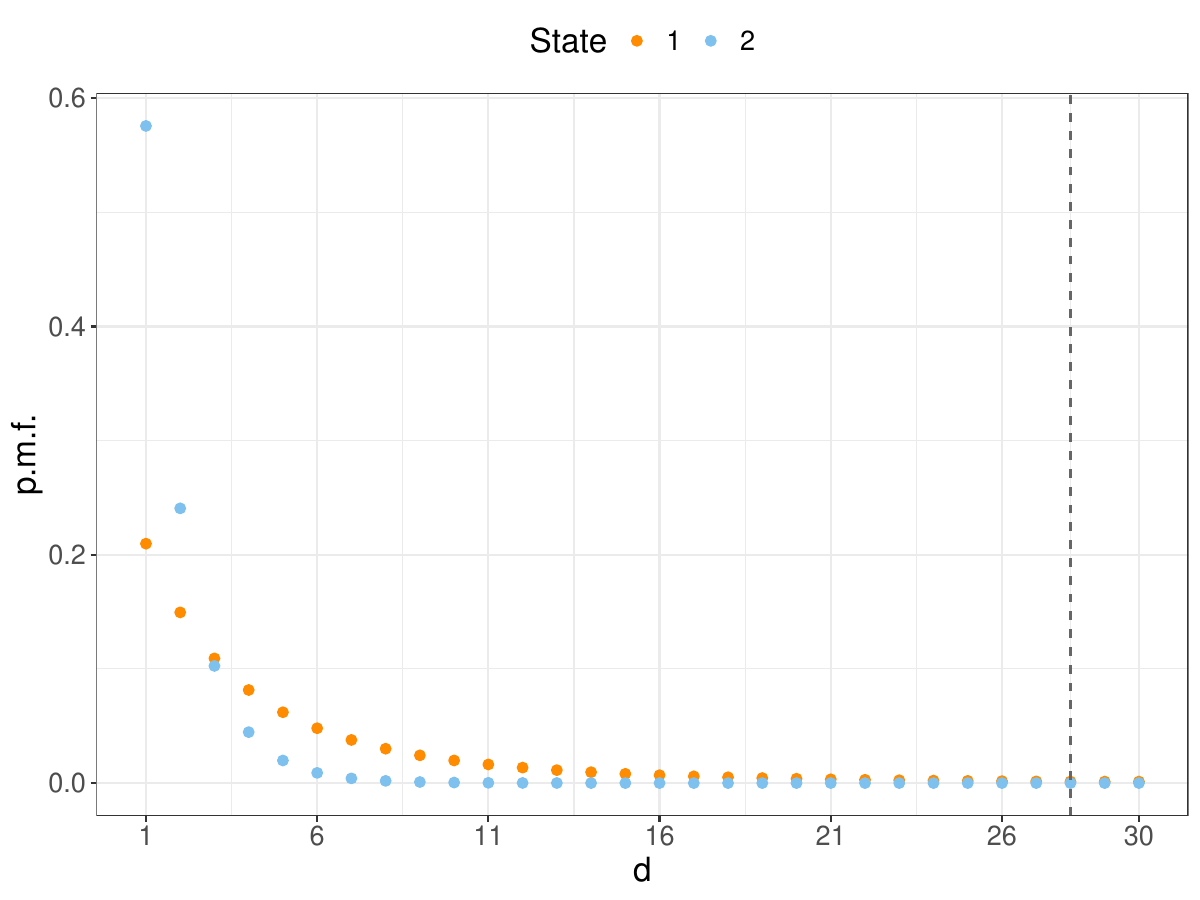}
    \caption{Estimated dwell time distribution for each state where the covariates are set to their median value. The dashed line indicates the value of $m=28$.}
    \label{fig:estdwelldistr}
\end{figure}

Figure \ref{fig:estdwelldistr} shows the estimated distributions of the dwell times across the two states $k=1,2$, when the covariates are set to their median values.
We see that the upper bound on the non-geometric dwell-time of $m=28$ is sufficiently large. As a matter of fact, both distributions seem to have an approximately geometric behavior after $d=21$.

\section*{Multivariate Risk Measures in the Environmental Context}
\begin{figure}
           \centering
           \includegraphics[width = .9\textwidth]{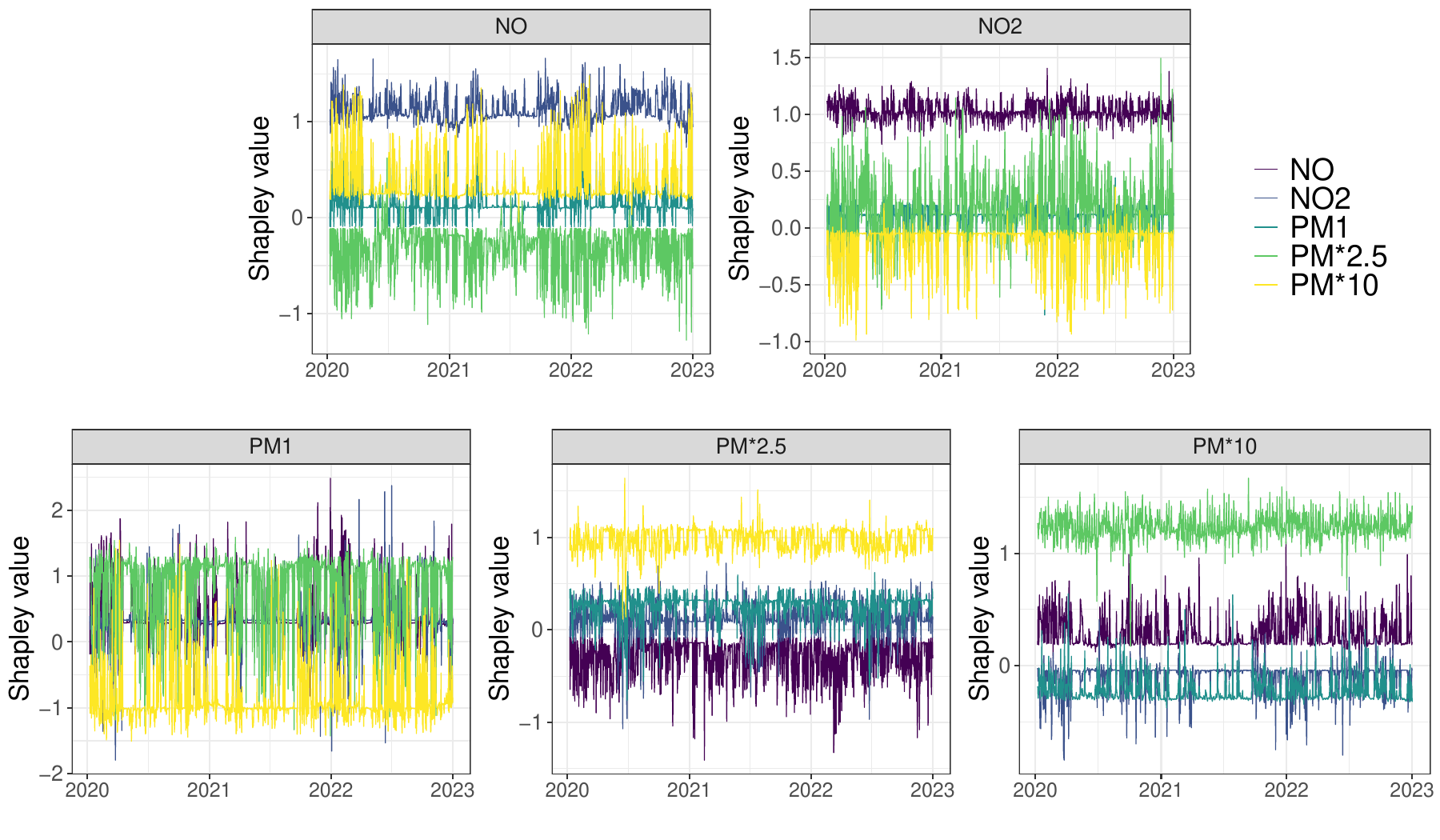}
           \caption{Time series of the Shapley Values of each pollutant on each other for the $MCoES$.}
           \label{fig:mcoes}
        \end{figure}

In the context of environmental risk, the risk is represented by the chances of observing a very large pollutant concentration. That is why the Value-at-Risk (VaR) of one pollutant $i\in\calS=\lcur 1,\dots,p\rcur$ can be defined as the minimum concentration that we would observe in the $\tau\times 100$ worst occasions and the Expected-Shortfall ($ES_i(\tau)$) as the tail-expectation given that the pollutant log-concentration is above the corresponding $VaR_i(\tau)$ value.
In practice, given that $Y_i\sim F_{Y_i}(y)=P(Y_i\leq y)$, we have that:
\begin{equation*}
    \begin{aligned}
    &VaR_{Y_{i}}(\tau)=F_{Y_{i}}^{-1}\lrnd 1-\tau\rrnd,\quad ES_{Y_{i}}(\tau)=\bbE\lsq Y_{i}\given Y_{i}\geq VaR_{Y_{i}}(\tau)\rsq.
    \end{aligned}
\end{equation*}
When dealing with multiple measurements observed jointly, say $\by\in\bbR^p$, these concepts must be extended to a multivariate setting.
\cite{tobias2016covar} introduces the $MCoVaR$ and $MCoES$ as multivariate counterparts of the marginal $VaR$ and $ES$. Rather than considering the marginal distribution of each element $i$, the quantile and tail expectations are evaluated on its conditional distribution given a certain partition of the remaining elements into $\calH_d\subset\calS\setminus\lcur i\rcur$ and $\calH_n=\calS\setminus\lcur \calH_d, i\rcur$. The pollutants in $j\in\calH_d$ are assumed to be in an \textit{in-distress} situation, represented by setting their values at the corresponding marginal $VaR_j(\tau^*)$ value with $\tau^*$ small (e.g. $0.05$). The pollutants in $\calH_n$ are assumed to be in a \textit{non-in-distress} situation, which is obtained by setting their values at their marginal median, i.e. $VaR_j(0.5)$.
In order to define these two measures formally, we introduce two vector-valued functions that associate a certain subset $\calH\subset\calS$ with the corresponding vectors of marginal $VaR$s and $ES$s:
\begin{equation}
    \label{eq:varsess}
    \begin{aligned}
    \bnu_\tau(\calH)=\lsq VaR_{{j}}(\tau)\rsq_{j\in\calH},\qquad\beps_\tau\lrnd \calH\rrnd=\lsq ES_{{j}}(\tau)\rsq_{j\in\calH}
    \end{aligned}
\end{equation}
The $MCoVaR$ and $MCoES$ of the outcome $i$ at level $\tau$ given $\calH_d, \calH_n$ are defined as:
\begin{equation}
    \label{eq:mcovarcoes}
    \begin{aligned}
        &MCV_{i}(\tau\given\calH_d, \calH_n)=F_{Y_{i}}^{-1}\lrnd 1-\tau\given \by_{\calH_d}=\bnu_{\tau^*}\lrnd \calH_d\rrnd, \by_{\calH_n}=\bnu_{0.5}\lrnd \calH_n\rrnd\rrnd,\\
        &MCE_{i}(\tau\given\calH_d, \calH_n)=\bbE\lsq Y_{i}\given Y_{i}>VaR_{i}(\tau), \by_{\calH_d}=\bnu_{\tau^*}\lrnd \calH_d\rrnd, \by_{\calH_n}=\bnu_{0.5}\lrnd t,\calH_n\rrnd\rsq,
        \end{aligned}
\end{equation}
where $\by_\calH$ denotes the subvector of $\by$ with indices belonging to $\calH$.
Both these measures assumes different values according to the assumed partition into $\calH_d\cup\calH_n=\calS\setminus\lcur i\rcur$.
To assess the overall contribution of each pollutant in a multivariate risk setting, we consider using the \textit{Shapley Value methodology}. This approach has been developed in game theory, for fairly distributing gains among a group of players who collaborate. In our context, as in the financial risk one, it is used to decompose the total risk among different risk factors. The overall contribution of every pollutant $j$ to the risk of pollutant $i$ is marginalizing across all combinations of the other pollutants being in distress or not.
Let $\eta_{i}(\calH_d)$ denote any of the multivariate risk measures defined in Equation \eqref{eq:mcovarcoes} for pollutant $i$, given that the subset $\calH_d \subset \calS \setminus \lcur i \rcur$ is in distress while $\calS \setminus \calH_d$ is not.
The overall contribution to the risk of the $i$-th pollutant by the $j$-th one can be quantified as:
\begin{equation}
    \label{eq:shapley}
    Sh_{i}(j)=\sum_{\calH\subset\calS\setminus\lcur i, j\rcur}v(\calH)\cdot\lrnd\eta_{i}\lrnd\calH\cup\lcur j\rcur\rrnd-\eta_{i}\lrnd\calH\rrnd\rrnd,
\end{equation}
where $v(\calH)=\frac{|\calH|!\cdot \lrnd |\calS\setminus\lcur i\rcur|-|\calH|-1\rrnd!}{|\calS\setminus\lcur i\rcur|!}$. Each term quantifies the variation in the risk of pollutant $i$ when pollutant $j$ is in distress rather than not-in-distress and is averaged across all possible configurations of the other pollutant being in distress or not.
All the elaborations in Equations \eqref{eq:mcovarcoes} and \eqref{eq:shapley} can be easily extended to the temporal dynamic context simply by accounting for the temporal heterogeneity in the conditional distributions of each pollutant.
Figure \ref{fig:mcoes} reports the results of this dynamic Shapley value for each pollutant under the MCoES risk measure. Comments are equivalent to those referred to the MCoVaR in the Main text.

\end{document}